\documentstyle[12pt,epsfig]{article}
\def\reference{\parskip 0pt\par\noindent\hangindent 0.5 truecm}

\def\lbar{\lambda \!\!\! ^-}
\begin{document}
%
%
\title{Factors Determining Variability Time in Active Galactic Nucleus Jets}
%


\author{R.J. Protheroe$^{1}$ 
} 

\date{}
\maketitle

{\center
$^1$Department of Physics and Mathematical Physics,
The University of Adelaide, Adelaide, SA 5005, Australia\\rprother@physics.adelaide.edu.au\\[3mm]
}

%
\begin{abstract}
The relationship between observed variability time and emission
region geometry is explored for the case of emission by
relativistic jets.  The approximate formula for the jet-frame
size of the emission region, $R'=Dc\Delta t_{\rm obs}$ is shown
to lead to large systematic errors when used together with
observed luminosity and assumed or estimated Doppler factor $D$
to estimate the jet-frame photon energy density.  These results
have implications for AGN models in which low-energy photons are
targets for interaction of high energy particles and photons,
e.g.\ synchrotron-self Compton models and hadronic blazar models,
as well as models of intra-day variable sources in which the
photon energy density imposes a brightness temperature limit
through Compton scattering.

The actual relationship between emission region geometry and
observed variability is discussed for a variety of geometries
including cylinders, spheroids, bent, helical and conical jet
structures, and intrinsic variability models including shock
excitation.  The effects of time delays due to finite particle
acceleration and radiation time scales are also discussed.
\end{abstract}


\bigskip



\section{Introduction}

In the standard picture of active galactic nuclei (AGN),
accretion onto a super-massive black hole is via an accretion
disk, and a significant fraction of the accretion power (possibly
supplemented by tapping into the rotational energy of the black
hole) produces twin opposing relativistic jets moving outward
along the disk axis, with typical Lorentz factors $\Gamma \sim
2$--10 as inferred from very long baseline interferometry (VLBI)
observations.  The objects observed in high energy $\gamma$ rays
are ``blazars'', AGN in which one of the jets is closely aligned
toward the observer.  It is natural that in $\gamma$-rays we
should see preferentially those AGN with aligned jets because the
emission from the jet is Doppler boosted in energy and
relativistically beamed along the jet direction (for a discussion
of relativistic effects see Urry and Padovani 1995).  The
$\gamma$ ray emission from blazars is variable (as it is also at
optical, UV and X-ray energies).  Relativistic effects also cause
the observed variability time to be shorter than the time scale
over which the emission changes in the jet frame.

The spectral energy distribution (SED) of blazars shows two broad
peaks, the low energy peak extending from the infrared to the UV
or X-ray region of the spectrum, and the high energy peak
starting in the X-ray or $\gamma$ ray range.  The usual
interpretation is that relativistic electrons produce the low
energy part by synchrotron emission, and that the same electrons
produce the high energy part by Compton scattering the low energy
part and/or external photons to higher energies.  The 3rd EGRET
catalog of high-energy $\gamma$-ray sources (Hartman et al.\
1999) contains around 70 high confidence identifications of AGN,
and all appear to be blazars (Montigny et al.\ 1995, Mukherjee et
al.\ 1997).  Clearly, the $\gamma$-ray emission is associated
with AGN jets.  

Four BL Lac objects have been detected in the TeV energy range:
Mrk~421 (Punch et al.\ 1992), Mrk~501 (Quinn et al.\ 1996),
1E~S2344+514 (Catanese et al.\ 1998) and PKS~2155-304 (Chadwick
et al.\ 1999).  Recently, the spectrum of Mrk 501 has been
measured up to 24 TeV by the HEGRA telescopes (Konopelko et al.\
1999).  Several of the EGRET AGN show $\gamma$-ray variability
with time scales of $\sim 1$ day (Kniffen et al.\ 1993) at GeV
energies.  The TeV $\gamma$-ray emission of two BL Lacs shows
very rapid variability.  For Mrk 421, variability on a time scale
as short as $\sim 15$ minutes has been reported (Gaidos et al.\
1996).  In the case of Mrk 501, variability on a time scale of a
few hours was observed during the 1997 high level of activity,
and there is evidence of a 23 day periodicity (Protheroe et al.\
1998, Hayashida et al.\ 1998) interpreted in terms of a binary
black hole model for the central engine by Rieger and Mannheim
(2000).  These variability timescales place important constraints
on the models.  For example, the synchrotron self-Compton (SSC)
model appears to be just consistent with recent multi-wavelength
observations of Mrk~421 and Mrk~501 during flaring activity
(Bednarek and Protheroe 1997, 1999).  However, the allowed range
of physical parameters (Doppler factor and magnetic field) is
rather small, and this mechanism may well be excluded by future
observations.  For a recent review of TeV $\gamma$-ray astronomy
see Kifune (2002).

Rapid variability in intra-day variability (IDV) sources is a
long-standing problem as it implies apparent brightness
temperatures in the radio regime which may exceed 10$^{17}$ K or
relativistic beaming with extremely high Doppler factors,
coherent radiation mechanisms, or special geometric effects
(Wagner and Witzel 1995).  The very rapid flaring observed at TeV
and X--ray energies during flaring activity in blazars also
presents a challenge for any model and suggests a re-examination
of mechanisms which may cause very rapid variability would be
worthwhile.  In this paper I concentrate on how the observed
variability time is related to the geometry and motion of the
emission region, and thus to the photon energy density in the
emission region.  The blazar emission mechanisms to be discussed
include: a shock excited emission region, bent jets, a shock
propagating along a jet containing a helical structure and
illuminating parts of the helix by enhanced interactions/emission
of radiation such that the emission regions move along helical
paths, and highly oblique conical shocks in the jet.  Together
with geometry-specific time delays and variable Doppler boosting
associated with relativistic motion of the emission region along
a curved trajectory, it may well be possible to explain the
observed flaring activity and high brightness temperatures.
Another possibility briefly discussed in the context of bent jets
and conical shocks is that jets may be fueled on an irregular
time scale.

The observed variability time $\Delta t_{\rm obs}$, and some
assumed or estimated Doppler factor $D$ is often used to estimate
the jet-frame source radius, $R'\approx Dc\Delta t_{\rm obs}$.
The jet-frame photon energy density is then usually assumed to be
$U'_{\rm phot} \approx L'/4\pi {R'}^2 c$, where $L'$ is the
jet-frame bolometric luminosity given by $L'=D^{-4} 4\pi d_L^2 F$,
$F$ being the observed bolometric flux, and $d_L$ being the
luminosity distance.  However, this approach can lead to large
systematic errors in the jet-frame photon energy density.  This
is important because the energy density of photons of the low
energy part of the SED may determine the energy losses of
electrons, and the rate of up-scattering to $\gamma$-ray energies
in SSC models, and the rate of proton-photon collisions in
hadronic models (see, e.g., Mannheim and Biermann 1989, Protheroe
1997, Mannheim et al.\ 2001, M\"{u}cke and Protheroe 2001,
M\"{u}cke et al.\ 2002).  If the emission region is optically
thin, it may also have consequences for IDV sources as I will show
that it is quite possible that the photon energy density
responsible for the so-called Compton catastrophe may actually be
lower than usually estimated.  In the following sections I shall
discuss how the variability time is related to the emission
region geometry and intrinsic jet-frame variability, show how
this can lead to large systematic errors in the jet-frame photon
energy density, and discuss other scenarios which can lead to
rapid variability and high fluxes.


\section{Relationship between variability time and emission region geometry}

\begin{figure}[hbt]
\begin{center}
\epsfig{file=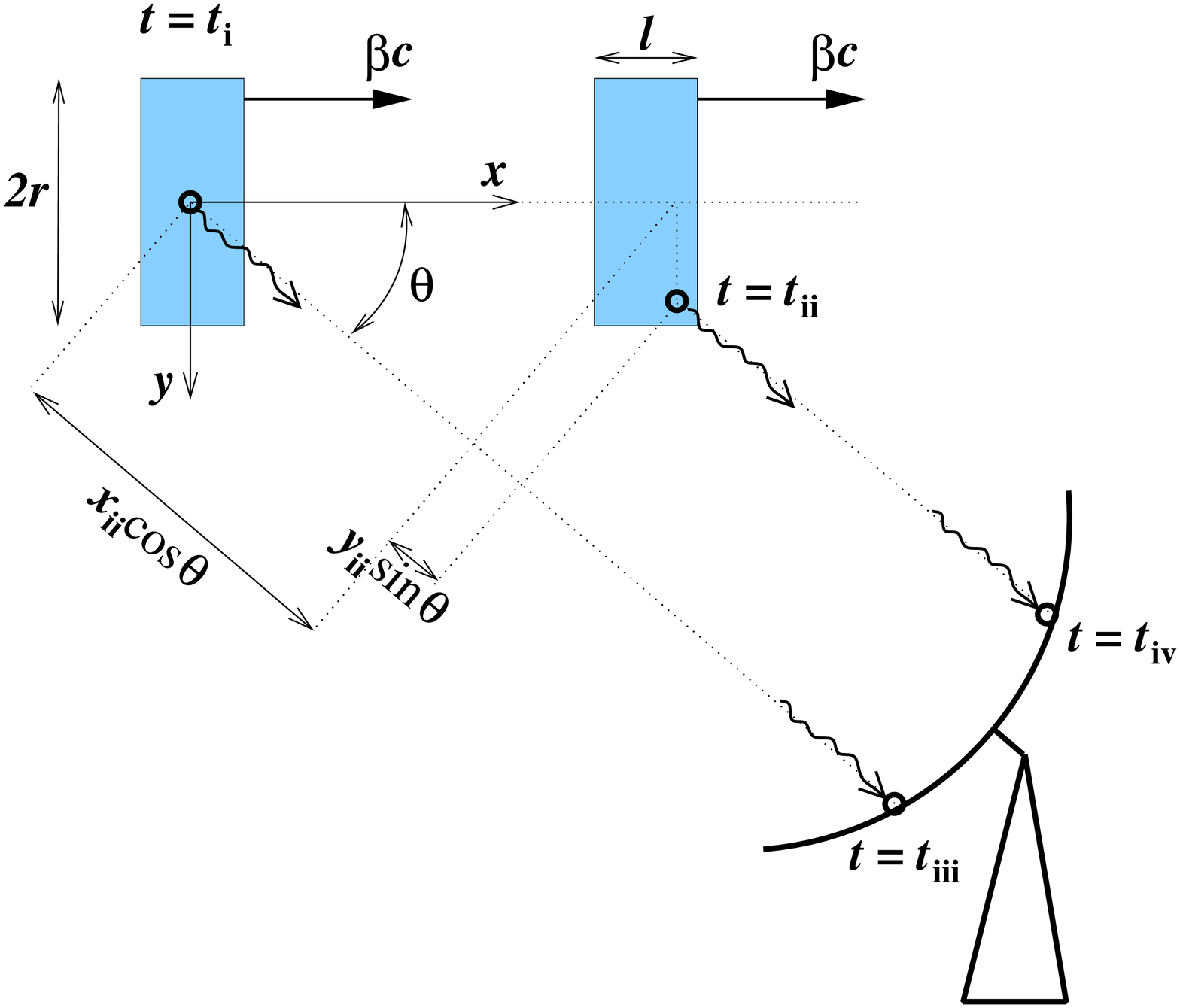, width=100mm}
\caption{Cylindrical emission region geometry. }
\label{geometry}
\end{center}
\end{figure}

I shall discuss first a relativistic jet pointing at angle
$\theta$ with respect to the line of sight to the observer, in
which the emission region is a cylinder of radius $r$ and
jet-frame length $\ell'$ moving along the jet with the jet's
Lorentz factor $\Gamma=1/\sqrt{1-\beta^2}$.  To work out how the
observed lightcurve depends on the emission region geometry and
the duration of the emission in the jet frame (primed
coordinates) we first consider in the observer frame (unprimed
coordinates) the events corresponding to: (i) the emission from
centre of cylinder at $(t\equiv t_{\rm i}=0,x\equiv x_{\rm i}=0,y\equiv y_{\rm i}=0,z\equiv z_{\rm i}=0)$ with this event
defining the origin of coordinates in both frames,
i.e. $(t'\equiv t'_{\rm i}=0,x'\equiv x'_{\rm i}=0,y'\equiv y'_{\rm i}=0,z'\equiv z'_{\rm i}=0)$ ; (ii) the emission from some
arbitrary point in the cylinder at a later time
$(t_{\rm ii},x_{\rm ii},y_{\rm ii},z_{\rm ii})$ ; (iii) the arrival at the
telescope, located in the $x$--$y$ plane, of the photon emitted
in event (i); (iv) the arrival at the telescope of the photon
emitted in event (ii) (see Fig.~\ref{geometry}).

In this paper, I shall assume that the distance between the AGN
and the observer is very much larger than the dimensions of the
emission region ($R$ and $\ell$ in Fig.~\ref{geometry}), and that
we are not concerned with investigating variability on time
scales very much shorter than $R/c$ or $\ell/c$.  Under this
assumption, the
time interval between arrival of the two photons at the telescope
is independent of distance to the AGN and is simply given by
\begin{eqnarray}
t_{\rm obs} \equiv (t_{\rm iv}-t_{\rm iii})= \left( t_{\rm ii}-{x_{\rm
ii}\over c} \cos\theta - {y_{\rm ii}\over c} \sin \theta \right).
\label{Eq:fundamental}
\end{eqnarray}
The error made by using this approximation is $\sim(R/d_L)(R/c)$
and is negligible compared with the variability time scale
associated with the emission region geometry investigated in this
paper which is $\sim(r/c)$.
 
Lorentz transformation to the observer frame,
gives $t_{\rm ii}=\Gamma(t'_{\rm ii}+\beta x'_{\rm ii})$, $x_{\rm ii}
=\Gamma(x'_{\rm ii}+\beta t'_{\rm ii})$ and $y_{\rm ii}=y'_{\rm ii}$. 
Noting that the Doppler factor is
$D= [\Gamma (1-\beta \cos\theta )]^{-1} = \Gamma (1+\beta
\cos\theta')$, and using the Aberration formulae,
\begin{eqnarray}
\cos\theta &=& {\Gamma \over D}(\cos\theta'+\beta), \;\;\; \sin\theta = {1\over D}\sin\theta'\\
\cos\theta' &=& {\Gamma  D}(\cos\theta-\beta), \;\;\; \sin\theta' = {D}\sin\theta,
\end{eqnarray}
Eq.~\ref{Eq:fundamental} becomes
\begin{eqnarray}
t_{\rm obs} &=& \Gamma(t'_{\rm ii}+\beta x'_{\rm ii}) -
\Gamma(x'_{\rm ii}+\beta t'_{\rm ii}){\Gamma \over Dc}(\cos\theta'+\beta) -
y'_{\rm ii}{1\over Dc}\sin\theta'\\
Dt_{\rm obs} &=& t'_{\rm ii}-{x'_{\rm
ii}\over c} \cos\theta' - {y'_{\rm ii}\over c} \sin \theta'.
\label{Eq:fundamental_jetframe}
\end{eqnarray}

Let us suppose that the cylindrical emission region emits 
radiation simultaneously and uniformly throughout its volume
between times $t'=-\Delta t'$ and $t'=\Delta t'$, as measured in
the jet frame.  While a simultaneous burst violates causality, it
is nevertheless a useful case to consider because it enables us
to determine clearly the contribution of emission region geometry
to the observed variability time.
For $\theta' \le \pi/2$ the first photon to arrive would have been emitted at
$(t'=-\Delta t', \, x'=\ell^{\,\prime}/2, \, y'=r, \, z'=0)$
giving
\begin{eqnarray}
t_{\rm obs}^{\rm first} = -\Delta t' - (\ell^{\,\prime}/2c)\cos\theta' -(r/c)\sin\theta'.
\label{Eq:fundamental_jet1}
\end{eqnarray}
Similarly, the last photon 
to arrive would have been emitted at
$(t'=\Delta t', \, x'=-\ell^{\,\prime}/2, \, y'=-r, \, z'=0)$
giving
\begin{eqnarray}
t_{\rm obs}^{\rm last} = \Delta t' + (\ell^{\,\prime}/2c)\cos\theta' +(r/c)\sin\theta'.
\label{Eq:fundamental_jet2}
\end{eqnarray}
For $\theta' > \pi/2$ the first photon to arrive would have been emitted at 
$(t'=-\Delta t', \,  x'=-\ell^{\,\prime}/2, \,  y'=r, \,  z'=0)$, 
and the last photon 
to arrive would have been emitted at 
$(t'=\Delta t',  \, x'=\ell^{\,\prime}/2, \,  y'=-r, \,  z'=0)$.
Hence, if we define the observer frame duration of the
burst as $2\Delta t_{\rm obs}$ then
\begin{eqnarray}
D \Delta t_{\rm obs} = \Delta t' + (\ell^{\,\prime}/2c)|\cos\theta'| +(r/c)\sin\theta'
\label{eq:tvarcyl}
\end{eqnarray}
and this is valid for all $\theta'$.  Note that Eq.~\ref{eq:tvarcyl} gives the usual
formula, $\Delta t_{\rm obs} = \Delta t'/ D$, if the emission
region is point-like (i.e. $\ell^{\,\prime}=r=0$).  If one term in
Eq.~\ref{eq:tvarcyl} dominates, $D\Delta t_{\rm obs}$ gives one
of: $\Delta t'$, $(\ell^{\,\prime} / 2c)|\cos \theta'|$, or $(r /c)\sin
\theta'$.

I shall next consider the case of a cylindrical emission region
in the jet being rapidly energized by a plane shock with
jet-frame speed $\beta'_{\rm shock}c$ travelling along the jet,
such that photons are emitted immediately after shock passage from a
thin disk-like region immediately downstream of the shock.  In
this case, the location of the emitting disk is defined by
$x'=\beta'_{\rm shock}ct'$, and so the arrival times of photons
at the telescope may be obtained from Eq.~\ref{Eq:fundamental_jetframe}
\begin{eqnarray}
cDt_{\rm obs} &=& x'\left({1\over \beta'_{\rm shock}} -\cos\theta'\right) - y' \sin \theta'.
\label{Eq:fundamental_jetshock}
\end{eqnarray}
For $({1/ \beta'_{\rm shock}} -\cos\theta') < 1$ the first and
last photons to be received would have been emitted at $(x'=\ell^{\,\prime}/2,
y'=r, z'=0)$ and $(x'=-\ell^{\,\prime}/2, y'=-r, z'=0)$, respectively.
However, for $({1/ \beta'_{\rm shock}} -\cos\theta') > 1$ the
first and last photons to be received would have been emitted at
$(x'=-\ell^{\,\prime}/2, y'=r, z'=0)$ and $(x'=\ell^{\,\prime}/2, y'=-r, z'=0)$,
respectively.  Hence, the observer frame duration of the burst is
\begin{eqnarray}
D \Delta t_{\rm obs} = (\ell^{\,\prime}/2c)\left|{1 \over \beta'_{\rm shock}}-\cos\theta'\right| +(r/c)\sin\theta'.
\label{eq:tvarcyl1}
\end{eqnarray}
If $\theta'$ is small ($\theta$ is {\em very} small), one finds
\begin{eqnarray}
 \Delta t_{\rm obs} & \approx
 &  { \ell^{\,\prime} \over 2D c}\left|{1-\beta'_{\rm shock}  \over \beta'_{\rm shock}}\right| + {r \theta' \over Dc}
\end{eqnarray}
and for a reasonable shock speed, e.g. $\beta'_{\rm shock}\sim \pm 0.5$,
\begin{eqnarray}
 \Delta t_{\rm obs} &{ \sim} & (2 \mp 1) { \ell^{\,\prime} \over 2D c} + {r \theta' \over Dc}
\end{eqnarray}
where $\beta'_{\rm shock}\sim - 0.5$ corresponds to a reverse shock.

What we have learned from this discussion is that multiplying
$c\Delta t_{\rm obs}$ by the Doppler factor might give the
jet-frame intrinsic variability time or {\em one} of the
dimensions of the emission region (possibly multiplied by some
unknown factor). One is tempted to ask if ${R'}\approx Dc\Delta
t_{\rm obs}$ is the right dimension to put in $U'_{\rm phot} =
L'/4\pi {R'}^2 c$ in order to estimate the jet-frame photon
energy density.  We shall discuss this point further in
Section~4.


\section{Monte Carlo investigation of $c\Delta t_{\rm obs}$, size, and $D$}

The Monte Carlo method allows the accurate calculation of
expected light curves for any emission region geometry and
intrinsic source variability.  The emission region is modeled in
the jet frame, and is represented by $N$ ``particles'', each of
which emits precisely one ``photon''.  The number density of the
particles models the geometry of the emission region, and each
particle emits its photon at a time determined by the emission
region geometry and variability model.  The jet-frame 4-position
of each photon emission event $(t'_i, x'_i, y'_i, z'_i)$,
$i=1,2,\dots N$, is determined by model.  The emission region
moves in the $x$-direction with Lorentz factor $\Gamma$, and the
observer-frame 4-position of each photon emission event $(t_i,
x_i, y_i, z_i)$, $i=1,2,\dots N$, is obtained by Lorentz
transformation.  The arrival time of each photon is calculated
for a given viewing angle $\theta$ using
Eq.~\ref{Eq:fundamental}, and is binned to give the lightcurve.

Two emission region geometries are considered: a solid cylinder,
and a spheroidal 3D gaussian.  We also consider four intrinsic
jet-frame time distributions: a simultaneous burst (violates
causality), a simultaneous gaussian pulse (violates causality)
and simultaneous emission at plane shock.  I shall investigate
effects of varying the Doppler factor $D$, viewing angle
$\theta$, jet Lorentz factor $\Gamma$, and shock speed
$\beta_{\rm shock}$.  Finally, the effect of smoothing due to
acceleration/radiation time delays is discussed.


\subsection{Solid cylinder emission region geometry}

\begin{figure}[hbt]
\centerline{\epsfig{file=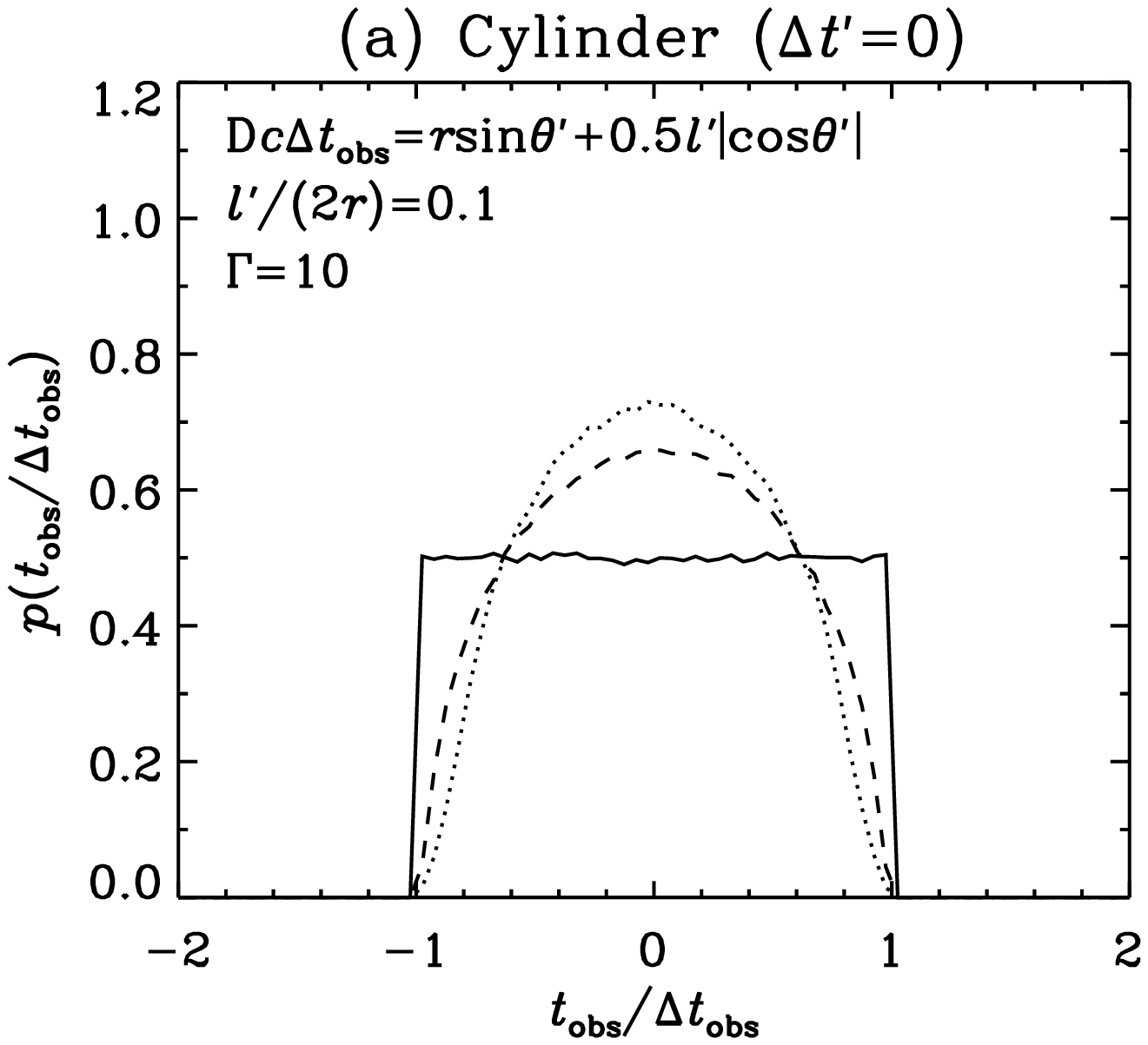, width=100mm}\hspace*{-15mm}\epsfig{file=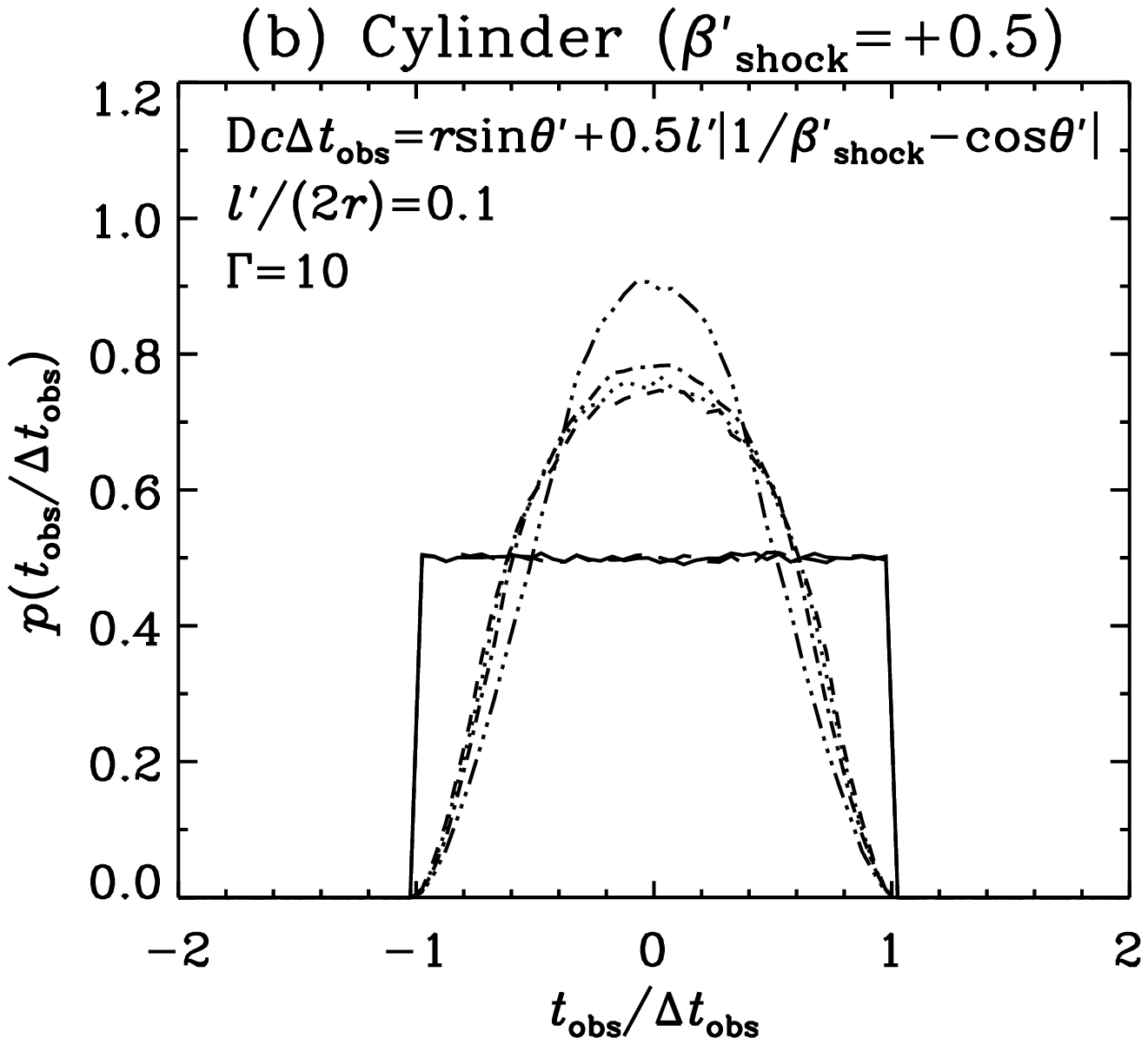, width=100mm}}
\caption{Lightcurve from cylindrical emission region with
$\ell/2r=0.1$ moving with Lorentz factor $\Gamma=10$ as observed
at various viewing angles: (a) Simultaneous burst for
$\theta'=0^\circ$ (solid), $36^\circ$ (dotted), $72^\circ$
(short-dashed), $108^\circ$ (short-dashed, identical to $72^\circ$),
$144^\circ$ (dotted, identical to $36^\circ$), $\theta'=360^\circ$
(long-dash, identical to $0^\circ$); (b) Excitation by a plane shock
moving at speed $\beta'_{\rm shock}c=0.5c$ for $\theta'=0^\circ$
(solid), $36^\circ$ (dotted), $72^\circ$ (short-dash),
$108^\circ$ (dot-dash), $144^\circ$ (dot-dot-dot-dash),
$\theta'=360^\circ$ (long-dash, identical to $0^\circ$).}
\label{agn_t_var_cylinder}
\end{figure}

If the emission region is ``solid'', i.e. the emissivity is
constant inside the emission region and zero outside, the
lightcurve may be peaked and have finite duration to reflect the
sharp edges of the emission region, and the shape of the
lightcurve will change with with viewing angle.  The Monte Carlo
results for the cylinder, when plotted such that the time of
observation is divided by the expected variability time $\Delta
t_{\rm obs}$ given by Eqs.~\ref{eq:tvarcyl}
and~\ref{eq:tvarcyl1}, is shown in
Fig.~\ref{agn_t_var_cylinder}(a) for a simultaneous burst
($\Delta t'=0$) , and in Fig.~\ref{agn_t_var_cylinder}(b) for
shock excitation.  As can be seen, Eqs.~\ref{eq:tvarcyl}
and~\ref{eq:tvarcyl1} are verified by the Monte Carlo results.
In both cases the shape of the lightcurve depends strongly on the
viewing angle.

\subsection{Gaussian emission region geometries }

The density of ``particles'' representing the emission region is described by 
\begin{equation}
\rho(x',y',z') \propto \exp\left[- \; \left({{x'}^2  \over 2 {\sigma_x}^2} + {{y'}^2 + {z'}^2  \over 2 {\sigma_r}^2}\right)\right]
\end{equation}
with $\sigma_x=\sigma_r$ corresponding to a spherical gaussian
distribution, $\sigma_x<\sigma_r$ to an oblate spheroidal
gaussian distribution, and $\sigma_x>\sigma_r$ to an prolate
spheroidal gaussian distribution.  For the case of a simultaneous
gaussian pulse, the probability of emission at jet-frame time
$t'$ to $(t'+dt')$ is $p(t')dt'$, where $p(t') = \exp(- {{t'}^2 /
2 {\sigma_t}^2})/\sqrt{2\pi}\sigma_t$ is independent of position.

If the emission region has a gaussian shape the lightcurve will
be smooth, and in many cases will also have a gaussian shape.
This is true, for example, for the cases of a simultaneous
gaussian pulse and excitation by a plane shock.  The width of the
lightcurve will depend on viewing angle in approximately the same
way as for equivalent cylindrical emission volume.  For example,
$r$ and $\ell'/2$ of the solid cylinder should be related to
$\sigma_r$ and $\sigma_x$, respectively, for the case of the
spheroidal gaussian density.  For the case of a simultaneous
gaussian burst I find the standard deviation to be given by
\begin{equation}
\sigma=[(\sigma_r\sin\theta'/c)^2+(\sigma_x\cos\theta'/c)^2+\sigma_t^2]^{1/2}/D.
\label{eq:gaussian_pulse}
\end{equation}
Note the similarity to Eq.~\ref{eq:tvarcyl} except that the terms
are added in quadrature as the standard deviation is required
instead of the maximum duration of the pulse which is, theoretically, infinite.

\begin{figure}[hbt]
\begin{center}
\epsfig{file=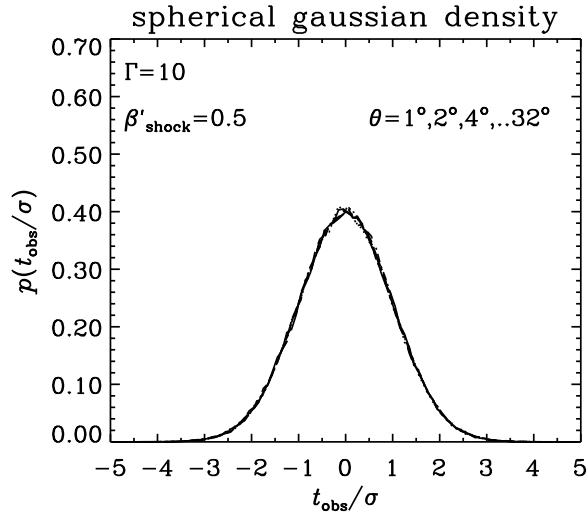, width=100mm}
\caption{Lightcurve due to excitation by a plane shock of a
spherical density moving with Lorentz factor $\Gamma=10$ for
$\beta'_{\rm shock}=0.5$ and various viewing angles shown is seen
to be a normal distribution.}
\label{agn_t_var_shpere_shock_a}
\end{center}
\end{figure}

In the case of excitation by a plane shock wave with speed
$\beta_{\rm shock}c$, the emission is from the plane where the shock
cuts the spherical gaussian density.  The emission region then has a
surface density of emitting particles which is a two-dimensional
gaussian, and the lightcurve reflects this distribution, and so is also
gaussian.  The standard deviation of the gaussian lightcurve
depends on the viewing angle as a result of projection effects,
and I find the standard deviation to be 
\begin{equation}
\sigma={1 \over cD}\left[(\sigma_r\sin\theta')^2+(\sigma_x)^2\left({1 \over \beta'_{\rm shock}}-\cos\theta'\right)^2\right]^{1/2}
\label{eq:gaussian_shock}
\end{equation}
Lightcurves for $\Gamma=10$, $\beta'_{\rm shock}=0.5$ and various
viewing angles are plotted in Fig.~\ref{agn_t_var_shpere_shock_a}
and seen to lie on top of each other when plotted in units of
$\sigma$, and to be a normal distribution.
Equation~\ref{eq:gaussian_shock} is valid for both forward and reverse
shocks.  For reverse shocks, $\beta'_{\rm shock}$ would be
negative and give rise to a broader lightcurves (larger $\sigma$).

\subsection{Acceleration/radiation delays}

Instantaneous excitation by a shock wave is not a realistic
approximation unless the time-scales for particle acceleration
and radiation (cooling) are very short compared to the transit
time of the shock through the emission region.  If this is not
the case, the lightcurve would be broadened and smoothed to
reflect the time delays associated with particle acceleration and
radiation.  An equation of the form $dN'/dt'=Q'(t')-N'/t'_{\rm loss}$
might describe the time-evolution of the number of particles, $N'$, radiating
photons at the frequency corresponding to that observed, with $Q'(t')$ being the source term,
and $t'_{\rm loss}$ representing the time-scale for particle losses (or time-scale over
which the radiation is emitted).
Then a simple smoothing function of the form
\begin{equation}
\phi_{\rm del}(t'_{\rm del}) = \left\{\begin{array}{ll} 0 & ( t'_{\rm del}< 0)\\
{1 \over \Delta t'_{\rm gain}}\left[1 - \exp\left({-t'_{\rm del} \over t'_{\rm loss}}\right)\right] & (0 \le t'_{\rm del} < \Delta t'_{\rm gain})\\
{1 \over \Delta t'_{\rm gain}}\left[1 - \exp\left({-\Delta t'_{\rm gain} \over t'_{\rm loss}}\right)\right]\exp\left[{-(t'_{\rm del}-\Delta t'_{\rm gain}) \over t'_{\rm loss}}\right] & (\Delta t'_{\rm gain} \le t'_{\rm del})
\end{array} \right.
\label{eq:smoother}
\end{equation}
can be used for this purpose, where $t'_{\rm del}$ represents
the delay between shock passage and emission by the
radiating particles, $\Delta t'_{\rm gain}$ represents the 
the duration of the acceleration following shock passage.  In the case of a pre-existing
population of thermal electrons shock heating is essentially
instantaneous ($\Delta t'_{\rm gain}=0$), and $t'_{\rm loss}$ represents
the time-scale for cooling by thermal bremsstrahlung radiation.
Similarly, for a pre-existing population of relativistic
electrons, the passage of the shock will essentially instantly
increase the magnetic field ($\Delta t'_{\rm gain}=0$) and $t'_{\rm
loss}$ would represent the time-scale for energy losses by
synchrotron emission.  If particle acceleration is required then
$\Delta t'_{\rm gain}>0$.

\begin{figure}[hbt]
\begin{center}
\epsfig{file=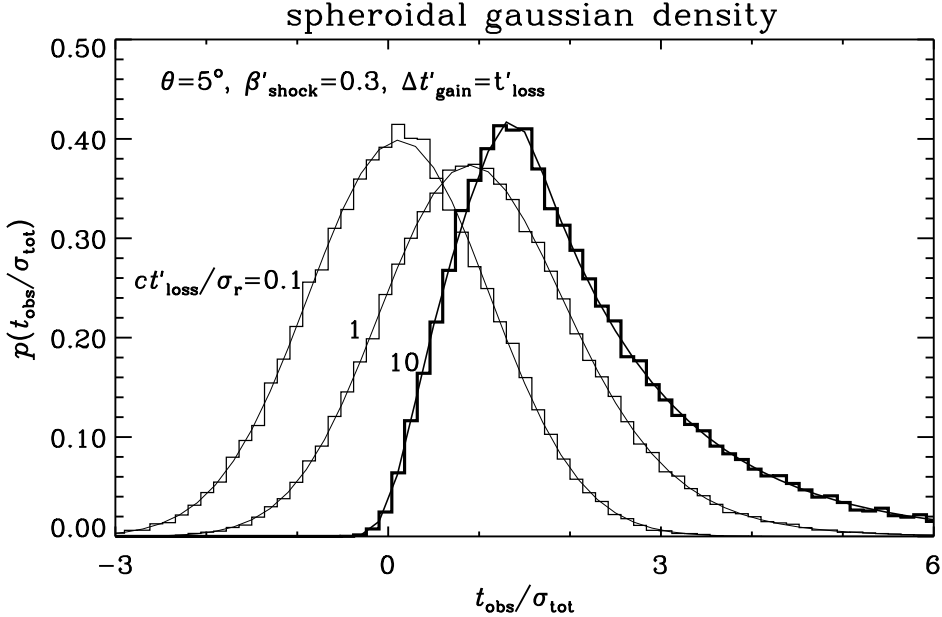, width=150mm}
\caption{Excitation of an oblate spheroidal gaussian density
($\sigma_x/\sigma_r=0.2$) moving with Lorentz factor $\Gamma=10$
by a plane shock with jet-frame speed $\beta'_{\rm shock}c=0.3c$,
followed by particle acceleration and radiation represented by
Eq.~\protect\ref{eq:smoother} with $\Delta t'_{\rm gain}=t'_{\rm
loss}$ for $t'_{\rm loss}=0.1\sigma_r/c$, $\sigma_r/c$ and
$10\sigma_r/c$ (as indicated) for viewing angle $\theta=5^\circ$.
Solid curves give the result of convolution
(Eqn.~\protect\ref{eq:convolve}).}
\label{agn_t_var_spheroidal_shock_delay}
\end{center}
\end{figure}

Fig.~\ref{agn_t_var_spheroidal_shock_delay} shows the effect of
acceleration/radiation delays on the lightcurve due to a plane
shock ($\beta'_{\rm shock}=0.3$) exciting an oblate gaussian
spheroidal emission region ($\sigma_x/\sigma_r=0.2$) with
acceleration/radiation delays described by $\Delta t'_{\rm
gain}=t'_{\rm loss}$ for $t'_{\rm loss}=0.1\sigma_r/c$,
$\sigma_r/c$ and $10\sigma_r/c$, and viewed at angle $\theta'=5^\circ$.  In the case of no
acceleration/radiation delays, the lightcurves would simply be
normal distributions centred on $t_{\rm obs}=0$.  Results are
shown for the three $t'_{\rm loss}$-values, and $t_{\rm obs}$ is plotted in
terms of $\sigma_{\rm tot}$ defined by  
\begin{equation}
\sigma_{\rm tot}^2 = \sigma^2 + {{t'_{\rm loss}}^2 +{\Delta t'_{\rm gain}}^2 \over 4cD},
\end{equation}
where $\sigma$ is given by Eqn.~\ref{eq:gaussian_shock}, such
that $\sigma_{\rm tot}$ gives a crude measure of the expected
duration of the lightcurve.  These distributions can be
obtained simply by convolution of a normal distribution
with $\phi_{\rm del}$ (Eq.~\ref{eq:smoother}), taking account of the fact that
jet-frame times enter in Eq.~\ref{eq:smoother}, that for any
point co-moving with the jet $\Delta t_{\rm obs} = \Delta t'/D$,
and that $t_{\rm obs}$ is plotted in units of $\sigma_{\rm tot}$, 
\begin{equation}
p\left({t_{\rm obs} \over \sigma_{\rm tot}}\right)= {\sigma D
\over \sqrt{2\pi}}\int_{-\infty}^\infty \phi_{\rm del}(t_{\rm
obs}D-t)\exp\left({-t^2\sigma_{\rm tot}^2\over 2\sigma^2}\right)
dt.
\label{eq:convolve}
\end{equation}
The solid curves, which agree with the Monte Carlo results
(histograms), are obtained from Eqn.~\ref{eq:convolve}.  As we
see, if the time-scales for the acceleration/radiation process
($\Delta t'_{\rm gain}$ and $t'_{\rm loss}$) are much less than
the time-scale $\sigma'$ associated with the shock passage and
dimensions of the emission region then the light-curve will be
symmetrical, and in the case of a gausian spheroidal emission
region will be a gaussian distribution with standard deviation
$\sigma$ (leftmost histogram).  If the time-scales for the
acceleration/radiation process are much larger than $\sigma'$,
then the light-curve will reflect that of the
acceleration/radiation process, i.e.\ Eqn.~\ref{eq:smoother}
(rightmost histogram).  The middle histogram shows an
intermediate case.

\begin{figure}[hbt]
\begin{center}
\epsfig{file=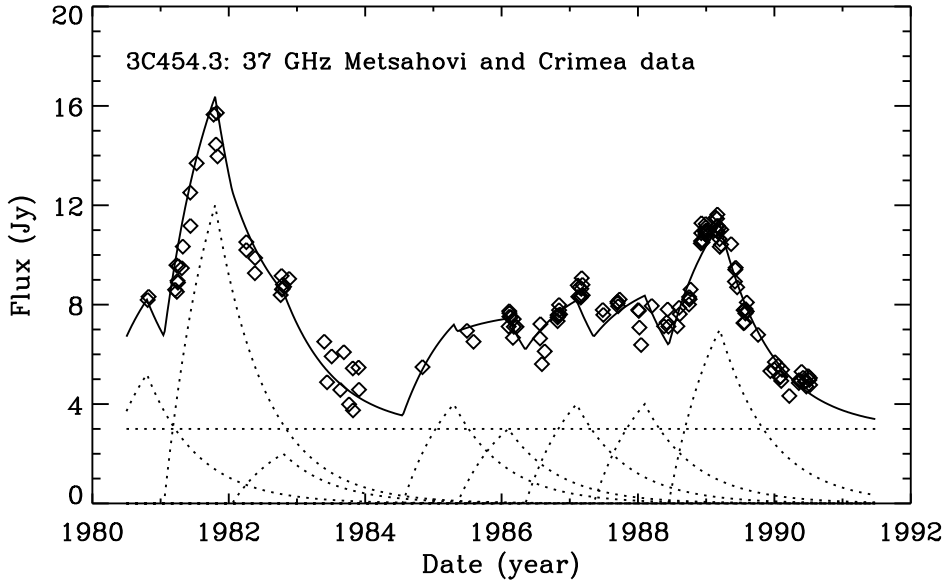, width=150mm}
\caption{Lightcurve of 3C~454.3 at 37~GHz (Salonen et al.\ 1987,
Terasranta et al.\ 1992).  The fit (solid curve) comprises a
background level (horizontal dotted line) and several flare
components (dotted curves) with individual flare lightcurves
given by Eq.~\protect\ref{eq:smoother} with $\Delta t_{\rm
gain}=t_{\rm loss}=0.75$~yr.}
\label{fig:example_3C454}
\end{center}
\end{figure}

Lightcurves of AGN show many different features and forms of
variability.  One example which appears to show variations
reflecting the acceleration/radiation process is 3C~454.3.  In
Fig.~\ref{fig:example_3C454}, I show the lightcurve of 3C~454.3
at 37~GHz obtained with the Metsahovi and Crimea telescopes over
ten years (Salonen et al.\ 1987, Terasranta et al.\ 1992).  The
flares appear non-symmetrical and have a shape similar to the
rightmost histogram in
Fig.~\ref{agn_t_var_spheroidal_shock_delay} ($\Delta t'_{\rm
gain}=t'_{\rm loss} \gg \sigma$).  As an example, I have
constructed a reasonably well-fitting lightcurve (solid curve)
from a number of flares of the form given by
Eqn.~\ref{eq:smoother} with $\Delta t_{\rm gain}=t_{\rm
loss}=0.75$~yr (dotted curves) plus a background flux density of
3~Jy.  One interpretation of these data would then be that the
radio emission region in 3C~454.3, being modelled by the solid
curve, has dimensions much less than $c$(0.75~yr)$D = 0.16D$~pc,
and that the energy-loss time-scale of the radiating electrons is
$\sim$(0.75~yr)$D$.

\begin{figure}[hbt]
\begin{center}
\epsfig{file=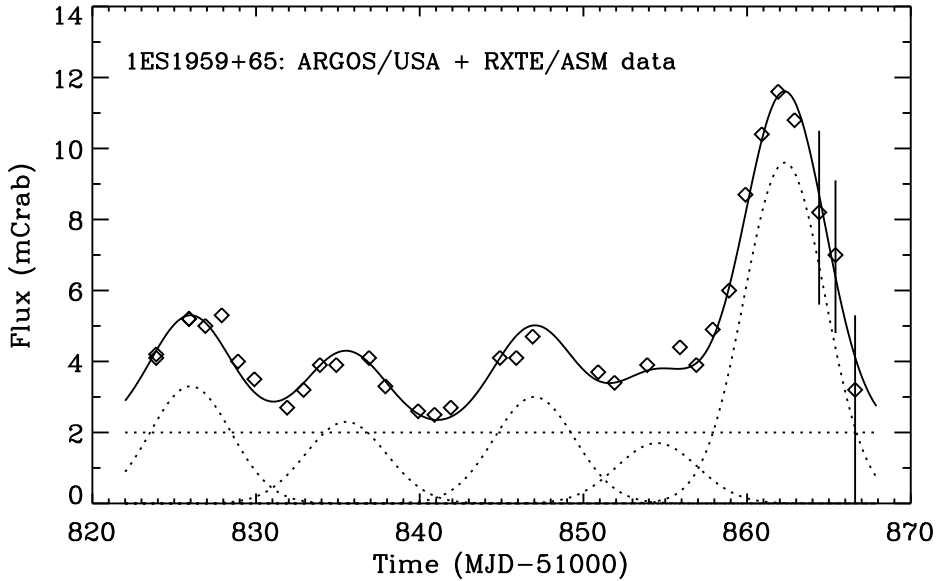, width=150mm}
\caption{X-ray lightcurve of 1ES1959+65 (Giebels et al.\ 2002).  The fit (solid curve) comprises a
background level (horizontal dotted line) and several flare
components (dotted curves) with individual gaussian lightcurves
with standard deviations $\sigma=2.5$~d.}
\label{fig:example_1ES1959}
\end{center}
\end{figure}

Another example is the X-ray lightcurve of 1ES1959+65 from
ARGOS/USA and RXTE/ASM (Giebels et al.\ 2002) shown in
Fig.~\ref{fig:example_1ES1959}.  In this case, the individual
flares appear to be symmetrical in time, and have a roughly
gaussian shape.  I have constructed a reasonably well-fitting
lightcurve (solid curve) from a number of flares of gaussian form
with standard deviations $\sigma=2.5$~d (dotted curves) plus a
background flux density of 2~mCrab.  One could interpret this as
indicating that the energy-loss time-scale of the radiating
electrons is much less than $2.5D$~d, and that the emission
region dimensions are $\sim c2.5D$~d$=2\times 10^{-3}D$~pc.


\section{Dependence of energy density on dimensions 
of the blob}

It is important to know the dimensions of the emission region for
several reasons: (i) in some hadronic models the synchrotron
photons are targets for photoproduction, (ii) in all models the
synchrotron photons are targets for photon-photon pair production
by $\gamma$-rays, (iii) in SSC models the synchrotron photon
energy density determines Compton scattering, and (iv) knowing
the systematic errors on photon energy density may help
understand the so-called ``Compton catastrophe'' in IDV sources
which have apparent brightness temperatures well in excess of
the limit $T_B < 10^{12}$~K imposed by Compton scattering
(Kellermann and Pauliny-Toth 1969, Kardashev 2000) when the
photon energy density in the emission region reaches the energy
density in the magnetic field.  One extreme example is
PKS~0405-385 (Kedziora-Chudczer et al.\ 1997) which has $T_B > 5
\times 10^{14}$~K after correcting for interstellar scintillation
(see also Walker 1998) requiring a Doppler factor of $D=10^3$ to
satisfy the brightness temperature limit.

To illustrate how critically the energy density depends on the
geometry, I shall consider the case of the jet-frame emissivity
following a spheroidal gaussian density.  Provided the emission
is optically thin, as is almost certainly true for the
optical--X-ray synchrotron hump in the SED of blazars, then it is
straightforward to calculate the average energy density from the
emission region geometry and the luminosity.

\begin{figure}[hbt]
\begin{center}
\epsfig{file=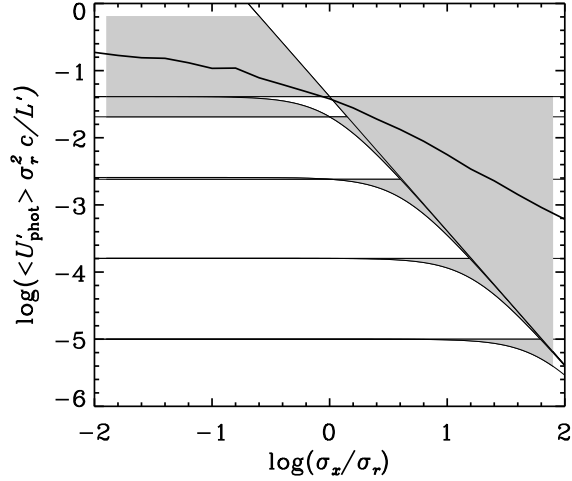, width=100mm}
\caption{Jet-frame average photon energy density versus
$\sigma_x$.  Thick solid curve shows actual density calculated
using the Monte Carlo method.  ``Observed'' density is shown by
the shaded areas for $\sigma_t=0$ (top shaded area), $\sigma_r$,
4$\sigma_r$, 8$\sigma_r$, and $16 \sigma_r$ (bottom shaded area).
See text for further details.}
\label{gaussian_spheroid_av_uphot}
\end{center}
\end{figure}

Assuming that the jet-frame luminosity, $L'$, is constant, we can
estimate the average jet-frame photon energy density given the
Doppler factor and emission region geometry.  The simplest way of
doing this, for any emission region geometry, is to use the Monte
Carlo method to place $N$ points at positions $\vec{r}_i, \;
i=1,\dots N$, distributed according to the emission region
geometry, and to give each point a luminosity $L'/N$.  Then at
point $\vec{r}_i$ the energy density is
\begin{eqnarray}
U'_{\rm phot}(\vec{r}_i) \approx {L' \over 4 \pi cN}\sum_{j \ne i} r_{ij}^{-2}
\end{eqnarray}
where $r_{ij} = |\vec{r}_i-\vec{r}_j|$.  Averaging over the
emission region distribution we obtain
\begin{eqnarray}
\langle U'_{\rm phot}\rangle \approx  {L' \over 4 \pi cN^2}\sum_{i=1}^N\sum_{j \ne i} r_{ij}^{-2}.
\end{eqnarray}
The result giving the jet-frame average photon energy
density versus $\sigma_x$ is shown as the thick solid curve in
Fig.~\ref{gaussian_spheroid_av_uphot} for the case of a gaussian
spheroidal density.

In Fig.~\ref{gaussian_spheroid_av_uphot} we also plot the
jet-frame average photon energy density that would be inferred if
we assumed that the observed variability time scale $\sigma$ and
an assumed or estimated Doppler factor $D$ gave the jet-frame
radius of a spherical emission region, i.e.\ $R'=\sigma D c$.
This ``observed'' jet-frame average photon energy density is
simply given by
\begin{eqnarray}
\langle U'_{\rm phot, \, obs}\rangle \equiv {L' \over 4 \pi (\sigma Dc)^2 c}.
\end{eqnarray}
Note that $\sigma$ (given by Eq.~\ref{eq:gaussian_pulse}) depends
on $\theta'$, $\sigma_t$ and $\sigma_x$ so that the ``observed''
jet-frame average photon energy density depends also on
$\theta'$, $\sigma_t$ and $\sigma_x$ as well as $D$.  The figure
shows $\langle U'_{\rm phot, \, obs}\rangle$ plotted against
$\sigma_x$ and gives the range due to variation in $\theta'$
(shaded) for various $\sigma_t$.  The upper bound in each case
gives the result for $\theta'=0$ or $\pi$ and the lower bound is
for $\theta'=\pi/2$.  We see that the ``observed'' value,
$\langle U'_{\rm phot, \, obs}\rangle$ can be several orders of
magnitude higher or lower than $\langle U'_{\rm phot}\rangle$ if
the emission region is different from a sphere, or if the
intrinsic variability time $\sigma_t$ is not small.  For example,
take the case of $\sigma_t \ll \sigma_r/c$, if $\sigma_x=10^{-2}
\sigma_r$ and $\theta'=0$ ($\theta=0$) then $\langle U'_{\rm
phot, \, obs}\rangle \approx 10^3 \langle U'_{\rm phot}\rangle$,
whereas if $\sigma_x=10^{2} \sigma_r$ and $\theta'=90^\circ$
($\theta=5.74^\circ$ for $\Gamma=10$) then $\langle U'_{\rm phot,
\, obs}\rangle \approx 10^2 \langle U'_{\rm phot}\rangle$. 
The above result has clear implications for both leptonic and
hadronic models of AGN in which photons of the low-energy peak of
the SED provide target photons for inverse-Compton scattering by
electrons (leptonic models) or pion photoproduction by protons
(hadronic models).  Using the observed variability time together
with assumed or estimated Doppler factor to estimate the emission
region radius $R'$ can clearly lead to large errors in $\langle
U'_{\rm phot}\rangle$.

Although the radio emission in IDV sources is usually assumed to
be optically thick, if this is not the case then the above result
may also have implications for IDV sources as the photon energy
density responsible for causing the brightness temperature limit
may actually be a few orders of magnitude lower than estimated on
the basis of time-variability, and in that case much lower
Doppler factors would be required to avoid the Compton
catastrophe.  I explore this further in a separate
paper (Protheroe 2002).


\section{Variability due to photon pile-up in observation time}

\begin{figure}[hbt]
\begin{center}
\epsfig{file=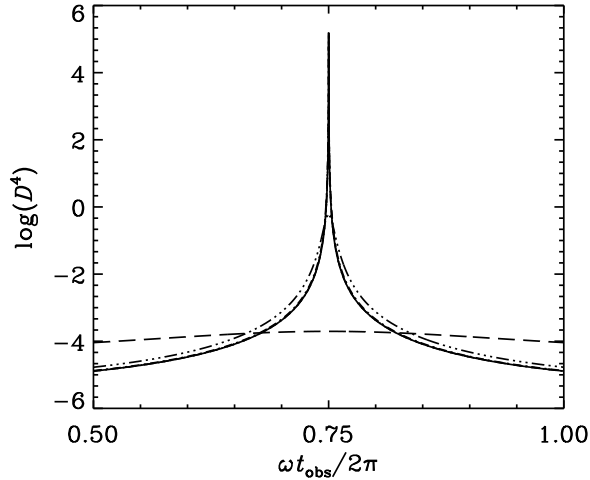, width=100mm}
\caption{Doppler factor raised to 4th power representing
bolometric lightcurve due to emission region moving along a bent
jet.  The jet Lorentz factor is $\Gamma=10$, the bent jet is
approximated by part of a circle around which jet plasma moves
with angular velocity $\omega$, and the emission region is
observed at angle $\eta \le 0.3^\circ$ (solid curve), $\eta
=27^\circ$ (chain curve), $\eta =81^\circ$ (dashed curve).}
\label{agn_bentjet}
\end{center}
\end{figure}

The simplest example of photon pile-up in observation time is a
bent jet.  Jets may bend if they pass through a stratified cold
and high density region (Mendoza and Longair 2001).  We
approximate the trajectory of an emission region moving with
speed $\beta c=(1-1/\Gamma^2)^{-1/2}$ along a bent section of jet
by motion around a section of a circle of radius $r$ in the
$x$--$y$ plane: $(t, \, x=r\cos(\omega t), \, y=r\sin(\omega t),
\, z=0)$, where $\omega=\beta c/r$.  For an observer in the
$x$--$z$ plane at angle $\eta$ to the $x$-axis, the observation
time is then given by
\begin{eqnarray}
t_{\rm obs} = t - (r/c) \cos (\omega t)\cos \eta
\end{eqnarray}
and the Doppler factor is 
\begin{eqnarray}
D = \{\Gamma [1 + \beta \cos \eta \sin  (\omega t)]\}^{-1}.
\end{eqnarray}
The Doppler factor raised to the 4th power is plotted against
observation time for $\Gamma=10$ and various observation angles
in Fig.~\ref{agn_bentjet}.  I have plotted $D^4$ as it is
appropriate for bolometric flux from a moving isotropic source;
it also applies to the specific flux $F_\nu$ for $F_\nu \propto
\nu^{-1}$.  We see that even for modest observation angles the
lightcurve is strongly peaked, essentially a delta function when
the emission region direction is closest to the line of sight.
Of course the finite size of any emission region will broaden the
distribution.  For example, for a lab-frame emission region length
along the jet $\ell$, the burst would have duration $\Delta
t_{\rm obs} \sim \ell/c$.

\subsection{Helical jet structures}

\begin{figure}[htb]
\begin{center}
\centerline{\epsfig{file=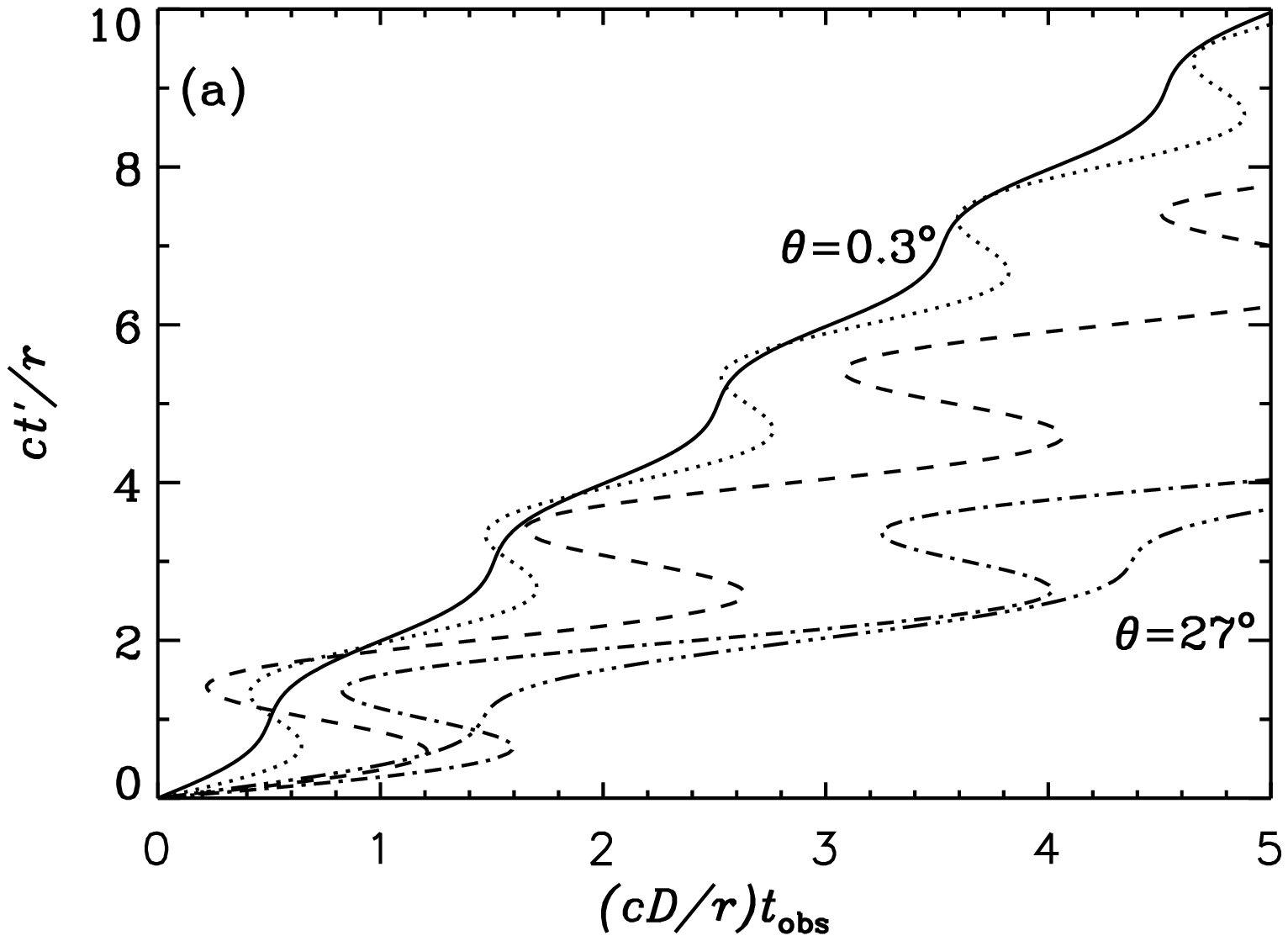, width=95mm}\hspace*{-10mm}\epsfig{file=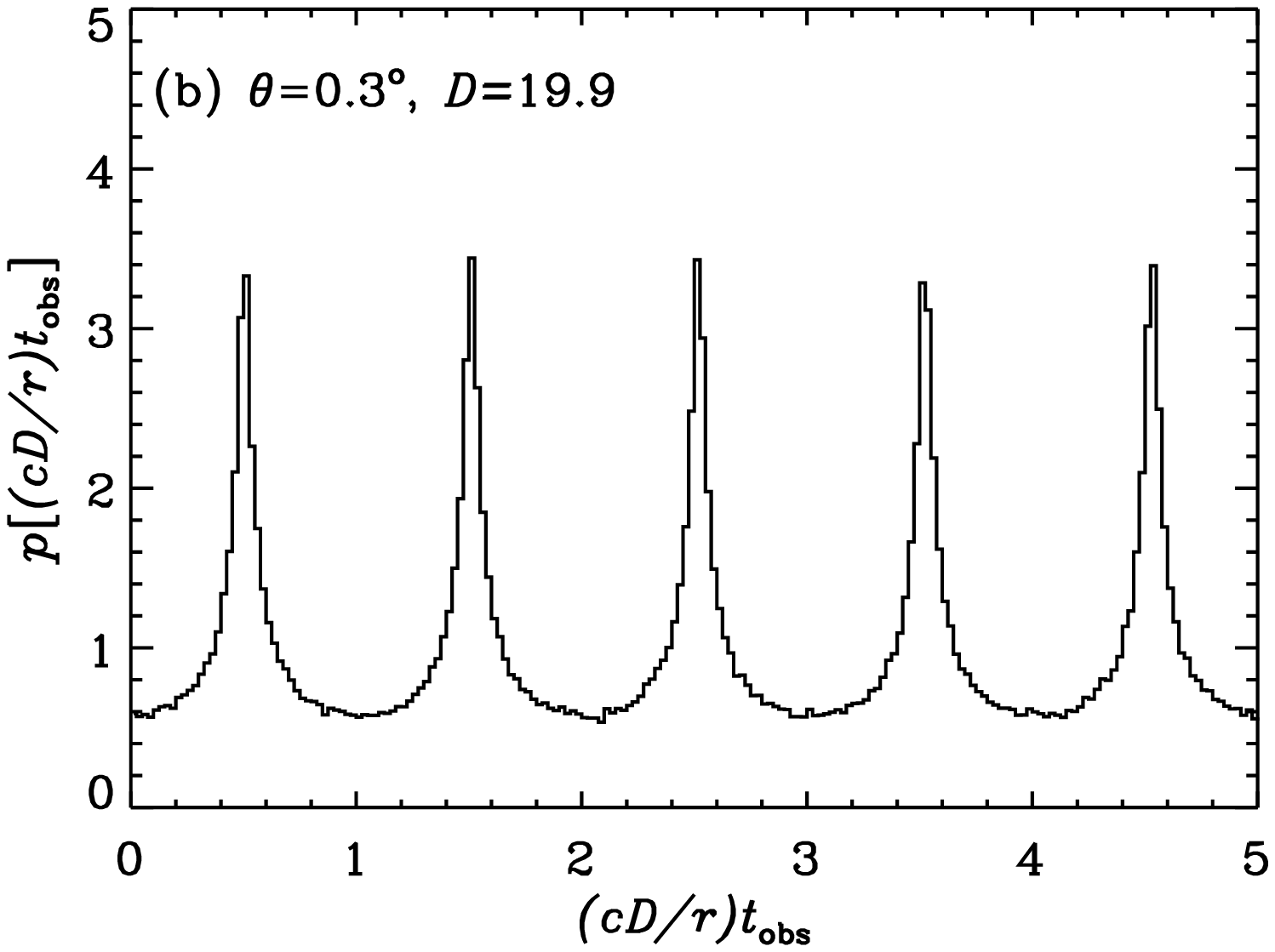, width=95mm}}
\centerline{\epsfig{file=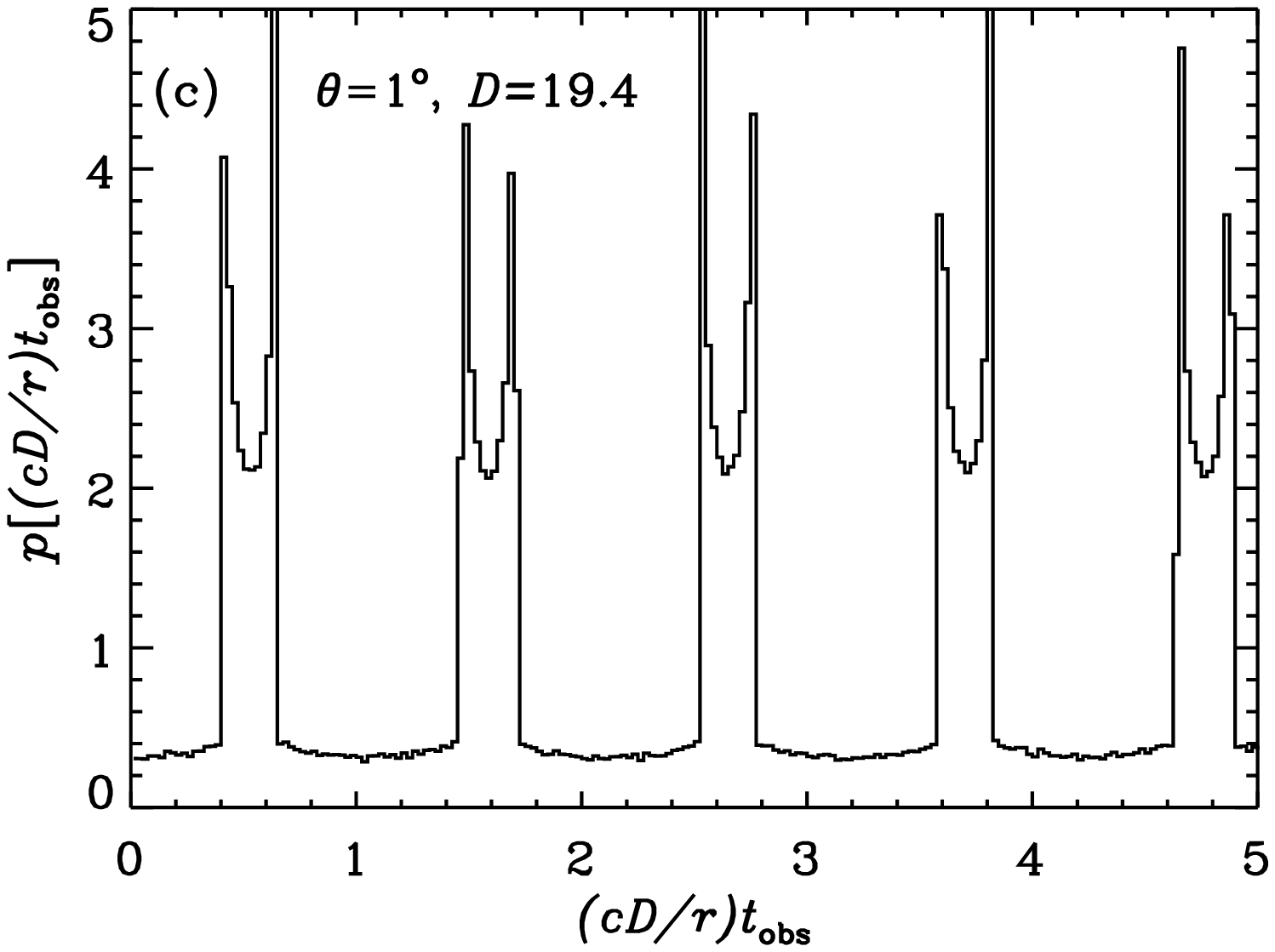, width=95mm}\hspace*{-10mm}\epsfig{file=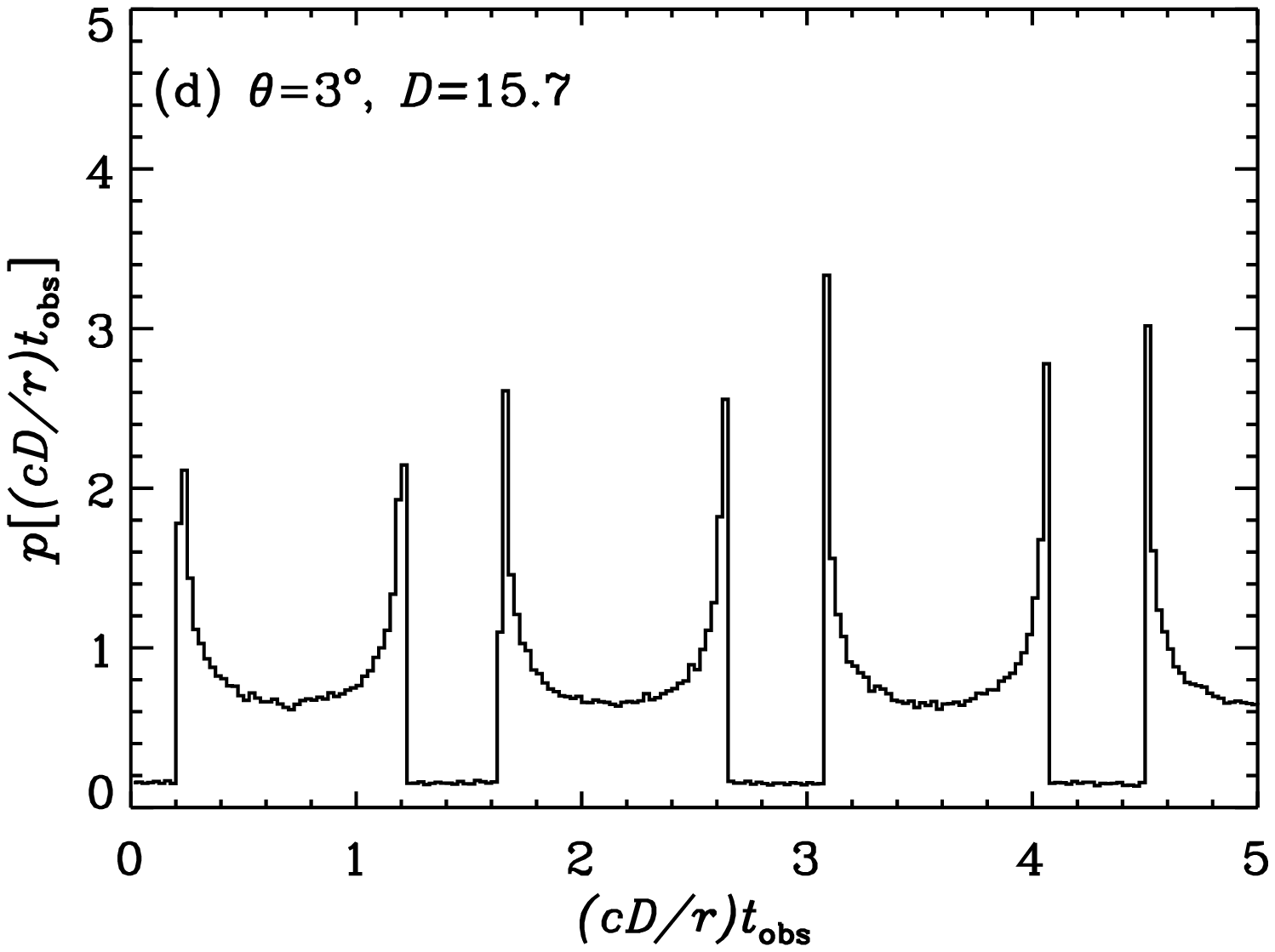, width=95mm}}
\centerline{\epsfig{file=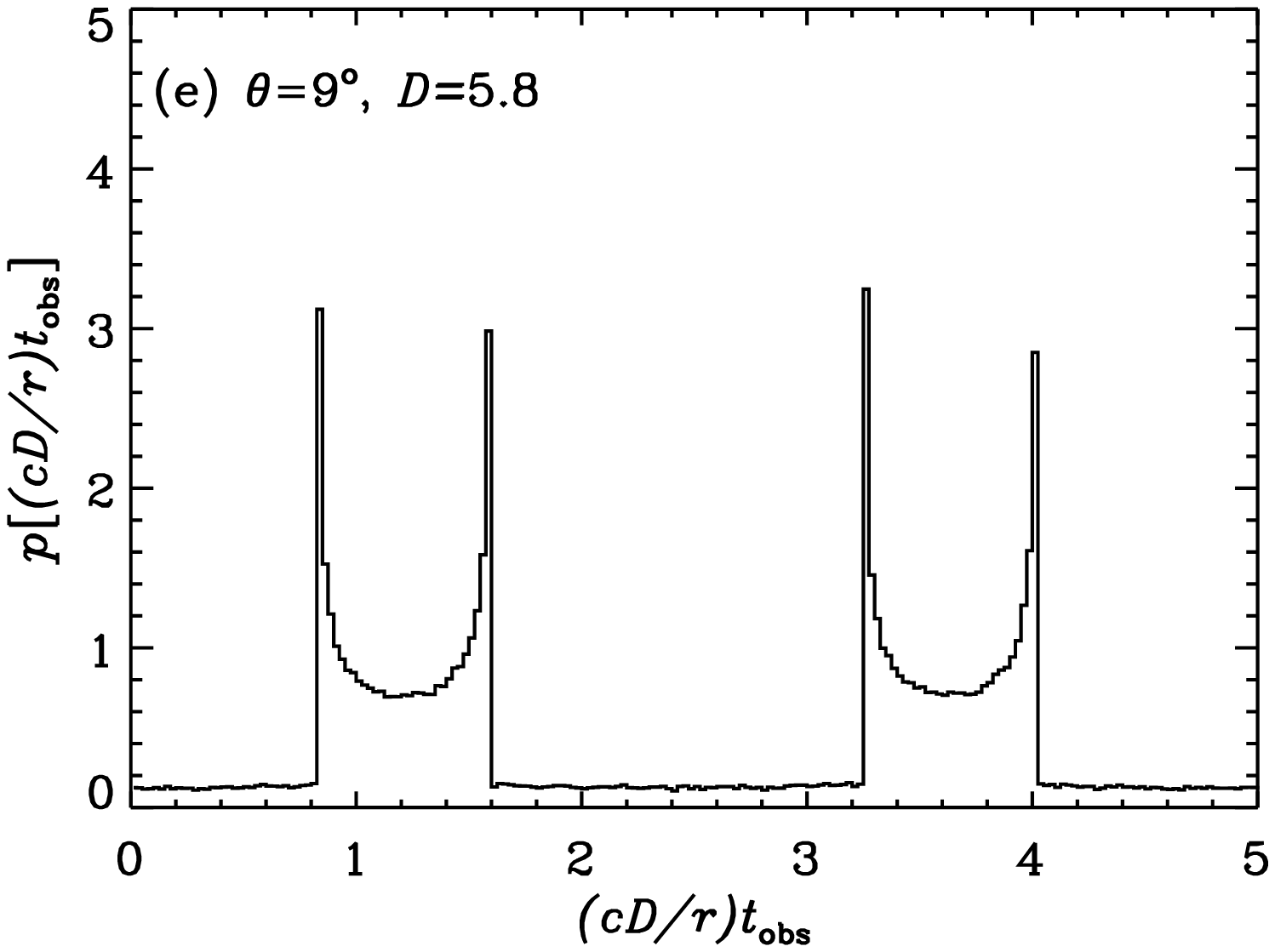, width=95mm}\hspace*{-10mm}\epsfig{file=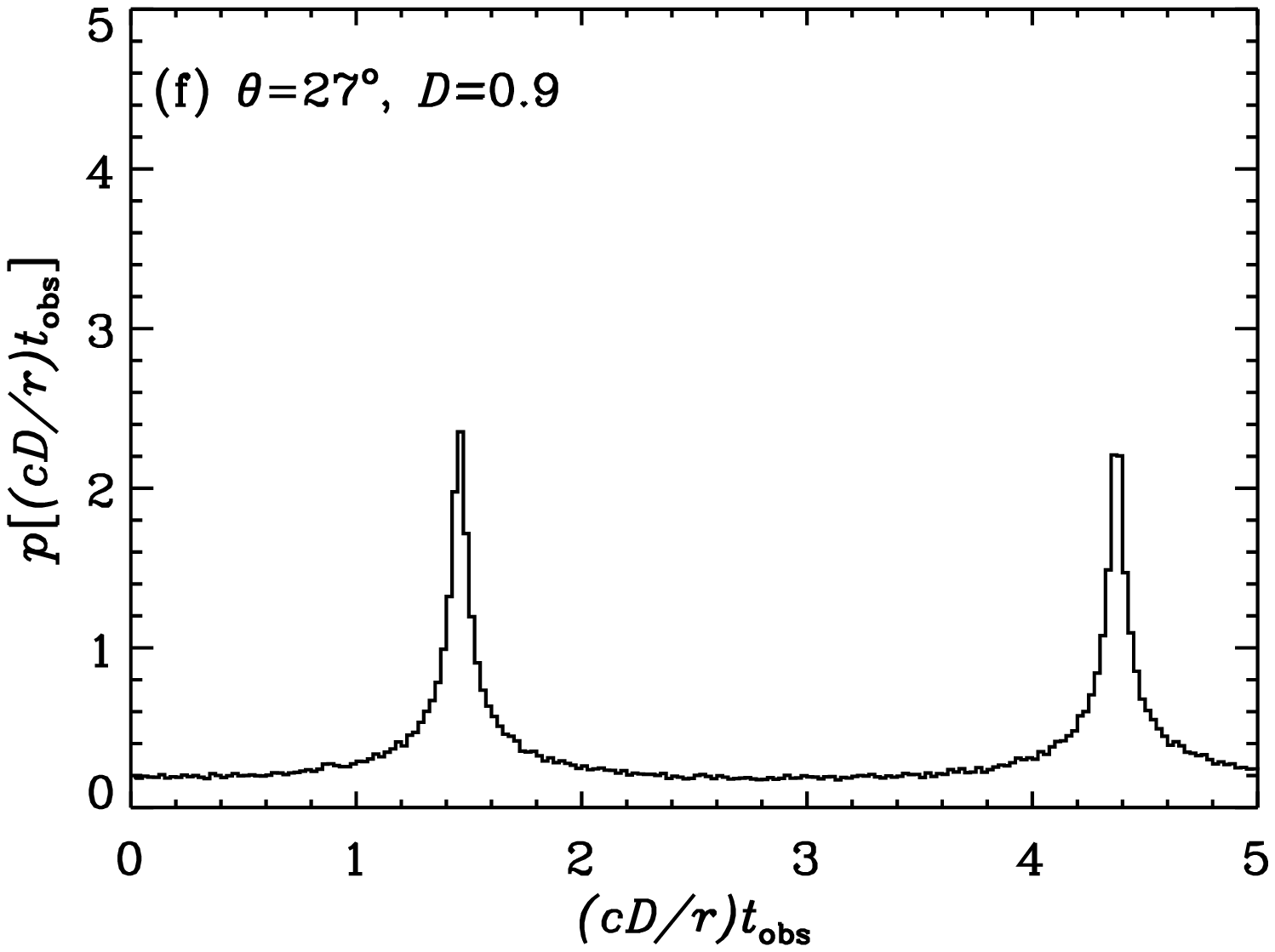, width=95mm}}
\caption{Shock excited helical jet structure for $\Gamma=10$, $\beta'_{\rm shock}=0.5$, $\lambda=r$.  (a) $t'$ vs.\ $t_{\rm obs}$ for  $\theta=0.3^\circ$ (solid curve), $1^\circ$ (dotted curve), $3^\circ$ (short dashed curve), $9^\circ$ (dot-dash curve), and $27^\circ$ (dot-dot-dot-dash curve).  Resulting lightcurves for (b) $\theta=0.3^\circ$, (c) $1^\circ$, (d) $3^\circ$, (e) $9^\circ$, and (f) $27^\circ$.}
\label{agn_tvar_helix}
\end{center}
\end{figure}

VLBI observations show that helical jets or helical structures in jets
may be fairly common in AGN (Rantakyro et al.\ 1998), and
theoretical studies have shown that wave-like helical structures
can occur as a result of jet precession (Hardee 2000).  Several
papers discuss helical jet models or the application of helical
models to specific sources (Camenzind 1986, Rosen 1990, Tateyama
et al.\ 1998, Qian et al.\ 1992, Schramm et al.\ 1993, Steffen et
al.\ 1995, Villata and Raiteri 1999).  Certainly helical jets or
structures would be important in determining the lightcurve of
$\gamma$-ray and neutrino emission from blazars, and various
suggestions have been made about the mechanisms involved
(Despringre and Fraix-Burnet 1997, Marcowith et al.\ 1995).

I consider a filamentary helical structure embedded in the jet
with Lorentz factor $\Gamma$ whose axis coincides with the jet
axis and is excited by a plane shock travelling along the jet
with jet-frame speed $\beta'_{\rm shock}c$.  The helical structure
could be, for example, a flux tube containing a relatively high
magnetic field, or a tube of high plasma density arising from a
density perturbation in the plasma entering the jet.  Helical
magnetic fields with an Archemedian spiral topology similar to the
``Parker spiral'' field of the heliosphere may well be expected
in AGN jets.

The excitation of a filamentary helical structure is described by
the jet-frame 4-vector
$(t',x',y',z')=(\lbar\phi/\beta'_{\rm shock}c, \,
\lbar\phi, \, r\cos\phi, \, r\sin\phi)$, where $r$ is the
radius of the cylinder containing the helix, $\lambda$ is the
helix wavelength, $\lbar=\lambda/2\pi$, $\phi=\beta'_{\rm
shock}ct'/\lbar$, and the jet is pointing in the
$x$-direction.  Lorentz transformation to the galaxy-frame gives
$(t,x,y,z)=[\Gamma t' (1+\beta\beta'_{\rm shock}), \,
\Gamma ct' (\beta+\beta'_{\rm shock}), \, r\cos(\beta'_{\rm
shock}ct'/\lbar), \, r\sin(\beta'_{\rm
shock}ct'/\lbar)]$.  The galaxy-frame speed of the location
of the excited part of the helix can exceed $c$, but this does
not violate causality as no particles or information
propagates at this pattern speed which is
\begin{eqnarray}
v_{\rm pattern} = {\beta'_{\rm shock} + \beta \over 1 + \beta'_{\rm shock}\beta}
\left[ 1 + \left( {r\beta'_{\rm shock} \over \lbar\Gamma(\beta'_{\rm shock} + \beta)}\right)\right]c.
\end{eqnarray}

Observation in the $x$-$z$ plane at angle $\theta$ to the jet
axis ($x$-axis) gives
\begin{eqnarray}
t_{\rm obs} &=& \Gamma t' [1 + \beta\beta'_{\rm shock} - (\beta'_{\rm shock}+\beta)\cos\theta]+r \sin\theta\sin(\beta'_{\rm shock}ct'/\lbar)/c \\
{dt_{\rm obs}\over dt'} &=& \Gamma [1 + \beta\beta'_{\rm shock} -\cos\theta(\beta'_{\rm shock}+\beta)]+r\sin\theta\,(\beta'_{\rm shock}/\lbar)\cos(\beta'_{\rm shock}ct'/\lbar).
\end{eqnarray}
Whenever $dt_{\rm obs}/ dt'=0$ the lightcurve will have a cusp.  If cusps are possible, they will occur at times corresponding to
\begin{eqnarray}
\cos\phi &=& {-\Gamma [1 + \beta\beta'_{\rm shock} -\cos\theta(\beta'_{\rm shock}+\beta)] \over r\sin\theta\,(\beta'_{\rm shock}/\lbar)}
\end{eqnarray}
provided the model parameters give $\cos\phi$ in the range $-1
\le \cos\phi \le 1$ ($\cos\phi$ depends on the helix geometry,
shock speed, jet Lorentz factor and viewing angle).  If
$\cos\phi=\pm 1$ one cusp per helix wavelength will occur, and if
$-1 < \cos\phi < 1$ multiple cusps occur, otherwise no cusps are
present in the light curve.  However, if $\cos\phi$ is close to
$\pm 1$ the lightcurve will be peaked.  This is illustrated in
Fig.~\ref{agn_tvar_helix} for $\Gamma=10$, $\beta'_{\rm
shock}=0.5$, and $\lambda=r$. Fig.~\ref{agn_tvar_helix}(a) shows
$t'$ plotted against $t_{\rm obs}$ for five viewing angles, and
Fig.~\ref{agn_tvar_helix}(b)--(f) shows the resulting lightcurve
for each of the five viewing angles (the corresponding Doppler
factors are also given).

Apart from the periodicity, the lightcurves shown in
Fig.~\ref{agn_tvar_helix} are reminiscent of those of blazars.
All of these lightcurves correspond to instantaneous emission
from the point on the helical filament at the time of excitation.
The cusps and peaks would in reality be smoothed to some extent
by the finite width of any helical structure as well as by any
delays associated with acceleration and radiation time scales.
Furthermore, as the shock weakens while propagating down the jet
the successive peaks/cusps in the lightcurve would decrease in
height.  One possible scenario could be that there is a
succession of shocks at random intervals propagate down the jet,
and because individual shocks weaken as they propagate, each shock
would cause only one or two peaks (due to only one or two cycles
of the helix).  In this case, it may not be possible to
distinguish between a simple bent structure in the jet and a
helical structure -- if appropriately aligned, a simple bent
structure would cause a cusp in the lightcurve in exactly the
same way as a helical structure.  An alternative to a helical
filamentary structure within the jet is a helical jet which would
produce a qualitatively-similar lightcurve, but would have the
Doppler factor varying with position along the jet as the viewing
angle relative to the local jet direction changes.


\subsection{Conical shocks}

Lind and Bladford (1985) have considered the
possibility that hotspots seen in VLBI images of radio jets may
actually be relativistically moving conical shocks.  They applied the
relativistic shock jump conditions to a plane parallel flow
entering a forward conical shock with cone angle $\eta$ to find
the angle $\zeta <\eta$ with at which the flow initially diverges
with respect to the jet axis.  Defining the downstream region
between the cones with angles $\zeta$ and $\eta$ to be the
emission region, and taking account of Doppler boosting, they
model the brightness distribution to simulate VLBI images.

\begin{figure}[hbt]
\begin{center}
\epsfig{file=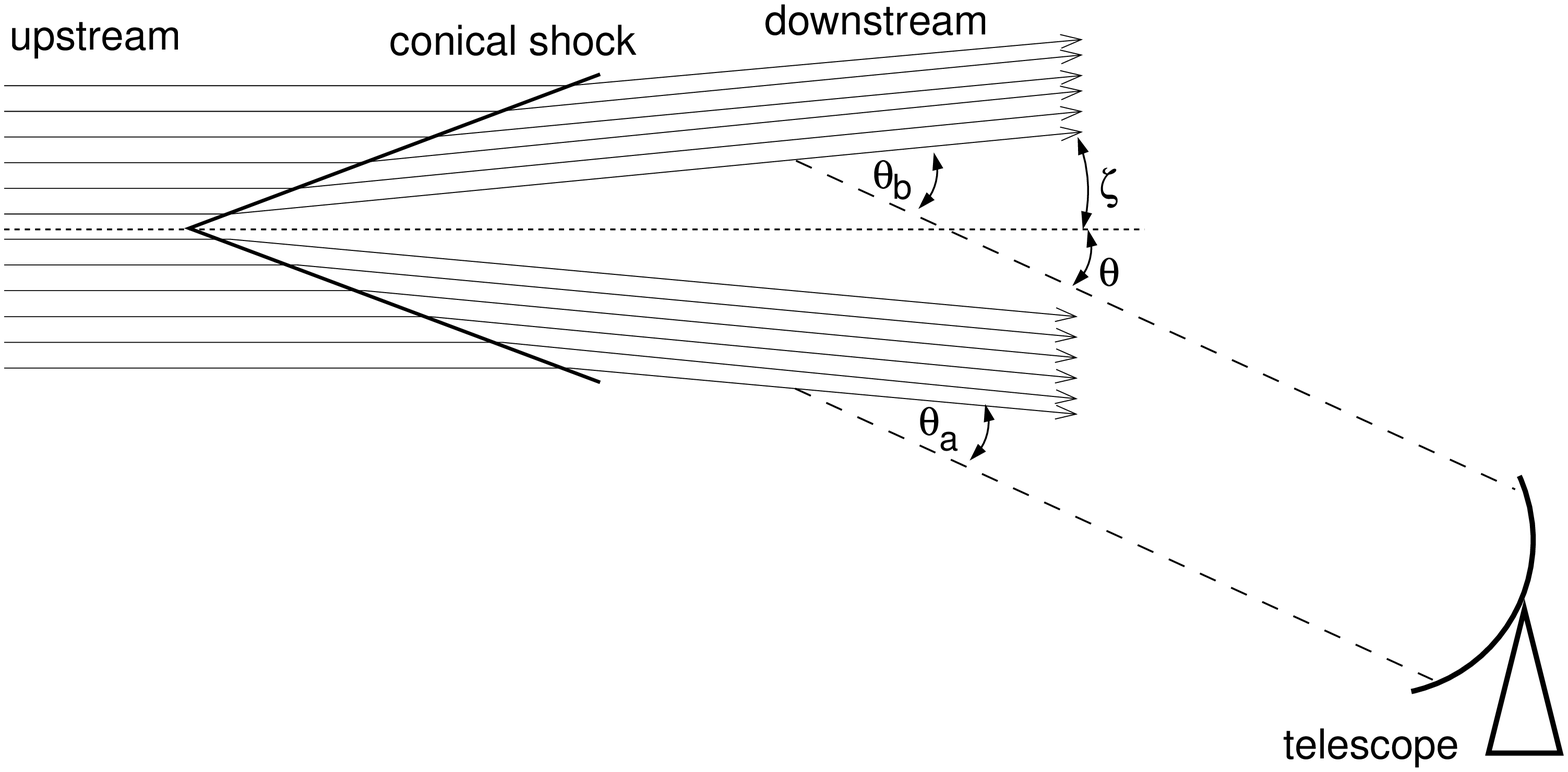, width=100mm}
\caption{Side view of the jet showing the location of a
stationary forward conical shock, and showing the downstream flow
diverging from the jet axis at angle $\zeta$ (not to scale).  For
a viewing angle $\theta$ with respect to the jet axis, the angle
between the downstream flow direction and the line of sight
varies between $\theta_a$ and $\theta_b$.}
\label{cone_doppler}
\end{center}
\end{figure}

The generation of conical shock structures is often seen to occur
in simulations of relativistic jets after the introduction of a
fast perturbation (Bowman 1994, Gomez et al.\ 1997).  These
structures, which are typically an alternating sequence of
forward and reverse conical shocks, are stationary or
slowly-moving in the galaxy frame, and have cone angles $\eta
\sim 1/\Gamma$.  Salvati et al.\ (1998) discuss emission by a
conical shock.  They consider the case of a density perturbation,
confined to thin flat disk, travelling relativistically along the
jet causing particle acceleration and emission where the disk
cuts a stationary forward conical shock such as that illustrated
in Fig.~\ref{cone_doppler}.

\begin{figure}[hbt]
\centerline{\epsfig{file=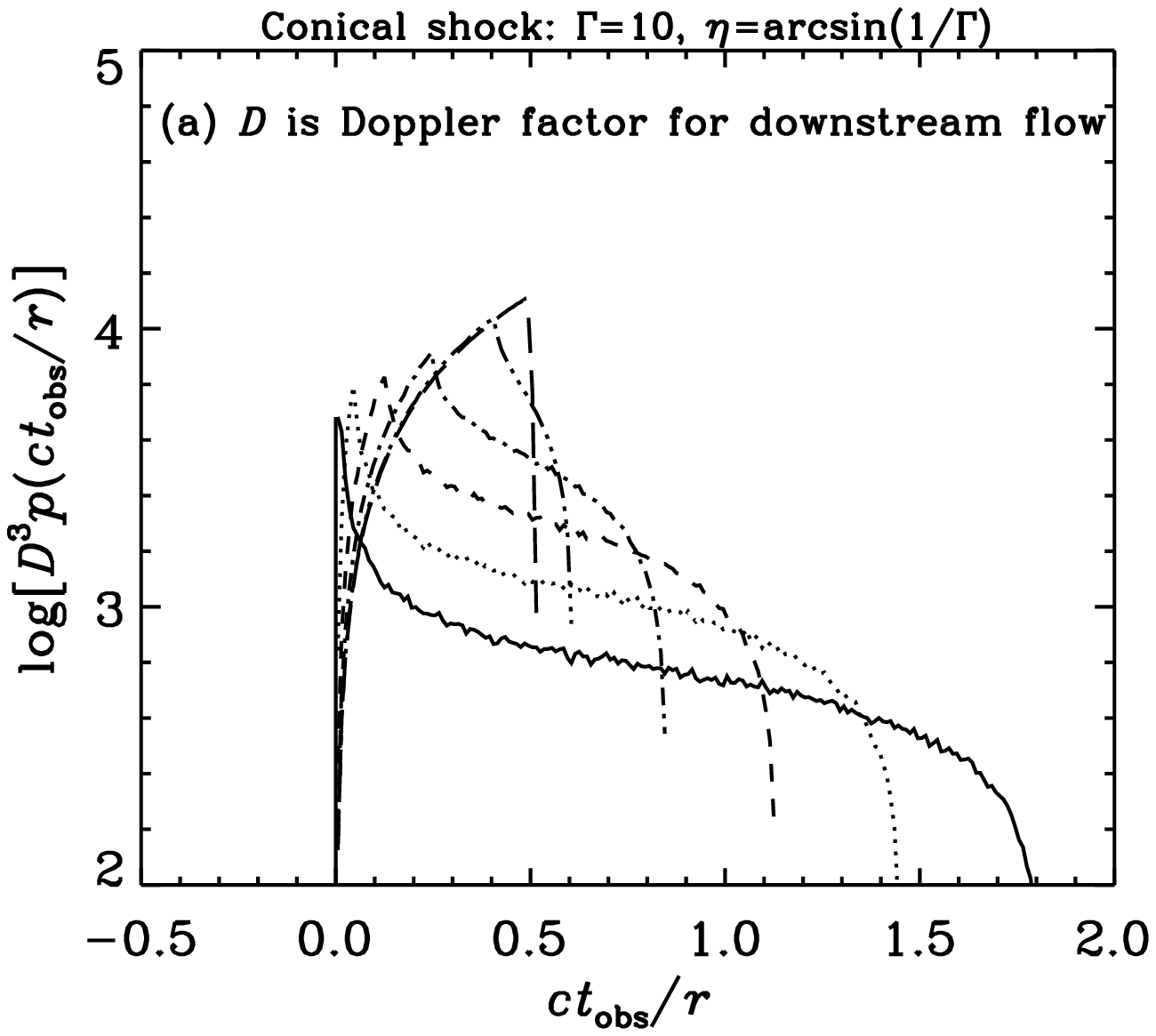, width=100mm}\hspace*{-15mm}\epsfig{file=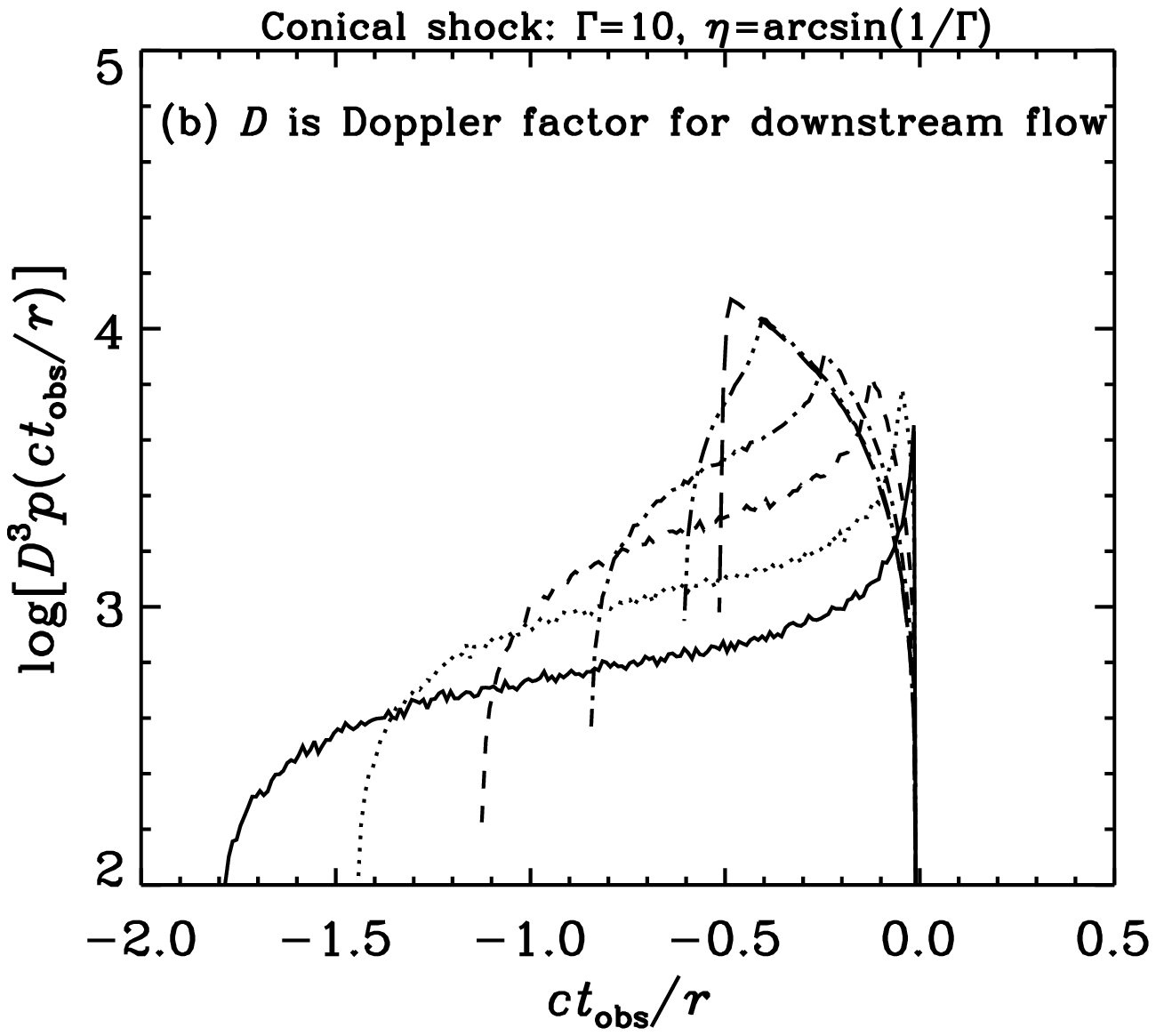, width=100mm}}
\caption{Lightcurve of (a) a stationary forward conical shock and
(b) a stationary reverse conical shock locally excited by a thin
density perturbation travelling along the jet with Lorentz factor
$\Gamma=10$ as viewed at angle $\theta$ with respect to the jet
axis.  The cone angle is $\eta=\sin^{-1}(1/\Gamma)$ and results are
given for $\theta/\eta=0.001$ (long dashed curves), 0.1
(dot-dot-dot-dash curves), 0.3 (dot-dash curves), 0.5 (short
dashed curves), 0.7 (dotted curves), 0.9 (solid curves).  The
Doppler factor is calculated using the local downstream flow
velocity.}
\label{agn_t_var_coneshock_boost_down}
\end{figure}

Salvati et al.\ (1998) assume that the time-scales for
acceleration and emission are negligible.  They show that their
model can lead to highly peaked light curves whose shape depends
on the viewing angle, and used these light curves to fit
Markarian 421 TeV flare data.  For the same input, and assuming
there is no Doppler boosting of the emitted radiation, I am able
to reproduce exactly their figure~2 which shows the observed flux
for various viewing angles.  However, bearing in mind that the
shock is stationary and that the pile-up in observing time is
already included, the bolometric flux emitted by part of the
downstream flow will be Doppler boosted by $D_{\rm local}^3$
where $D_{\rm local}=[\Gamma_d(1-\beta_d \cos\theta_{\rm
local})]^{-1}$ where $\Gamma_d$ is the Lorentz factor of the
downstream flow, and $\theta_{\rm local}$ is the viewing angle
with respect to the line of sight and the local downstream flow
direction which varies around the shock as indicated in
Fig.~\ref{cone_doppler} ($\theta_{\rm local}$ ranges between
$\theta_a$ and $\theta_b$).  To obtain the Lorentz factor of the
downstream flow, it is easiest to Lorentz transform in a
direction parallel to the shock plane to a frame in which the
flow is normal to the shock.  For the case of cone angle
$\eta=\sin^{-1}(1/\Gamma)$ assumed by Salvati et al.\ (1998),
i.e.\ an oblique shock at angle $\eta$ to the upstream flow, and
using the relativistic equation of state, I find that for
$\Gamma=10$, $\eta \approx 7.25^\circ$ and $\zeta \approx
1.89^\circ$ and that the Lorentz factor of the downstream flow is
related to that of the upstream flow by $\Gamma_d \approx
0.801784 \Gamma_u$.  My result for the observed flux for the same
input as Salvati et al., but including the Doppler boosting
taking into account the local downstream flow directions, is
given in Fig.~\ref{agn_t_var_coneshock_boost_down}(a) and shows
that the inclusion of Doppler boosting causes the peak at
$\theta=0$ to be higher than that at $\theta=0.9\eta$, the
opposite to that found by Salvati et al.  However, since Salvati
et al.\ (1998) used one viewing angle in their fits to Markarian
421 TeV flare data (their figure~3), and because the divergence
of the downstream flow is rather small, their fits are still
valid and their model remains an interesting mechanism for flare
production.  The same authors (Spada et al.\ 1999) have applied
their model to IDV sources and are able to explain brightness
temperatures up to $3\times 10^{17}$~K.

I wish to extend the work of Salvati et al.\ (1998) by including
the reverse shocks, and ultimately a sequence of stationary
reverse and forward shocks.  In
Fig.~\ref{agn_t_var_coneshock_boost_down}(b) I show the
lightcurve for a reverse conical shock having identical
parameters as the forward shock already discussed.  Because the
divergence of the downstream flow is rather small, the lightcurve
of the reverse shock is approximately just the lightcurve of the forward
shock reflected about $t_{\rm obs}=0$.

\begin{figure}[hbt]
\begin{center}
\epsfig{file=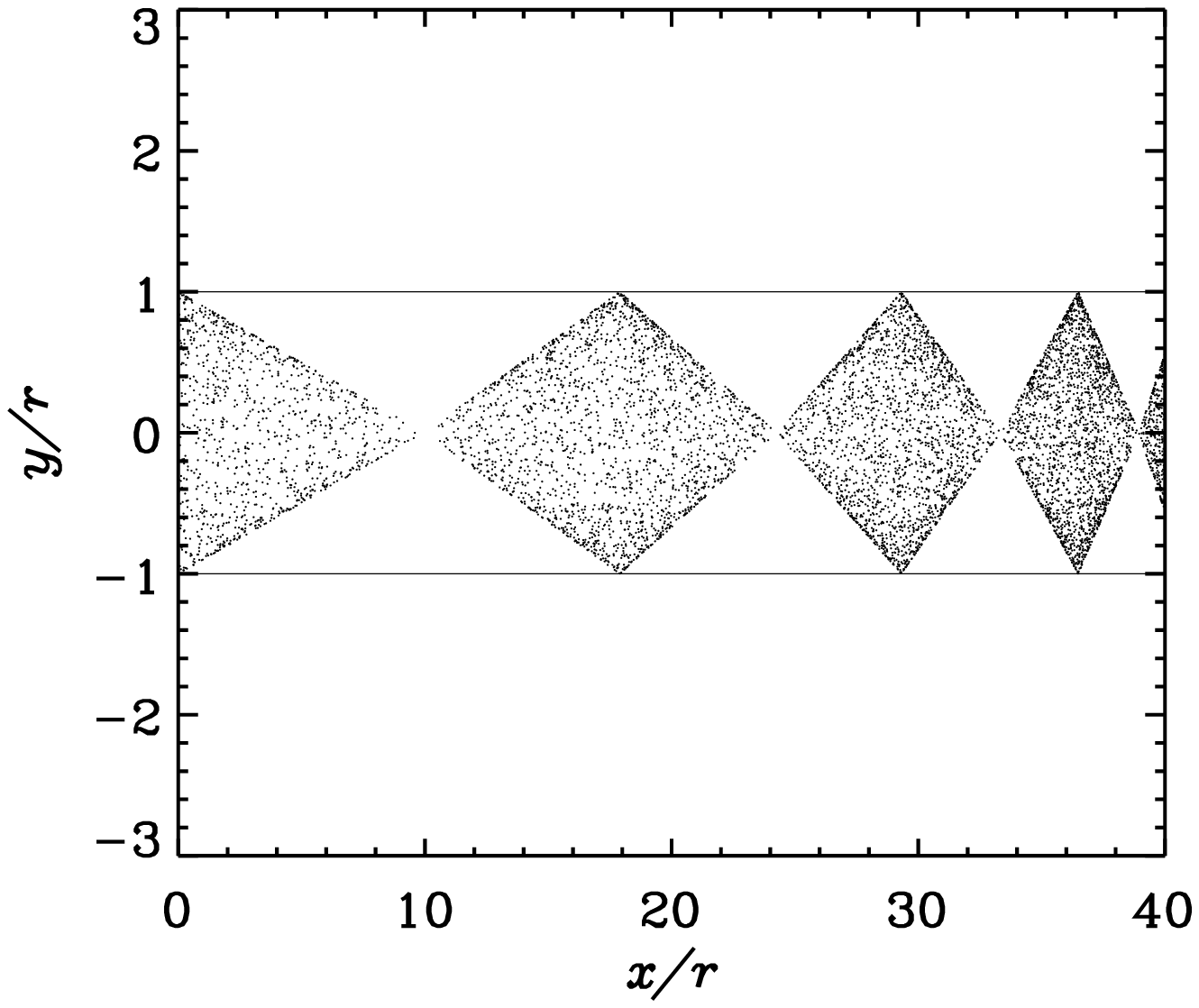, width=100mm}
\caption{Side view of the jet showing the location of
``particles'' used in the Monte Carlo Method.  The location of
the conical shocks is clearly evident.}
\label{LC_of_coneshocks_geom}
\end{center}
\end{figure}

\begin{figure}[hbt]
\centerline{\epsfig{file=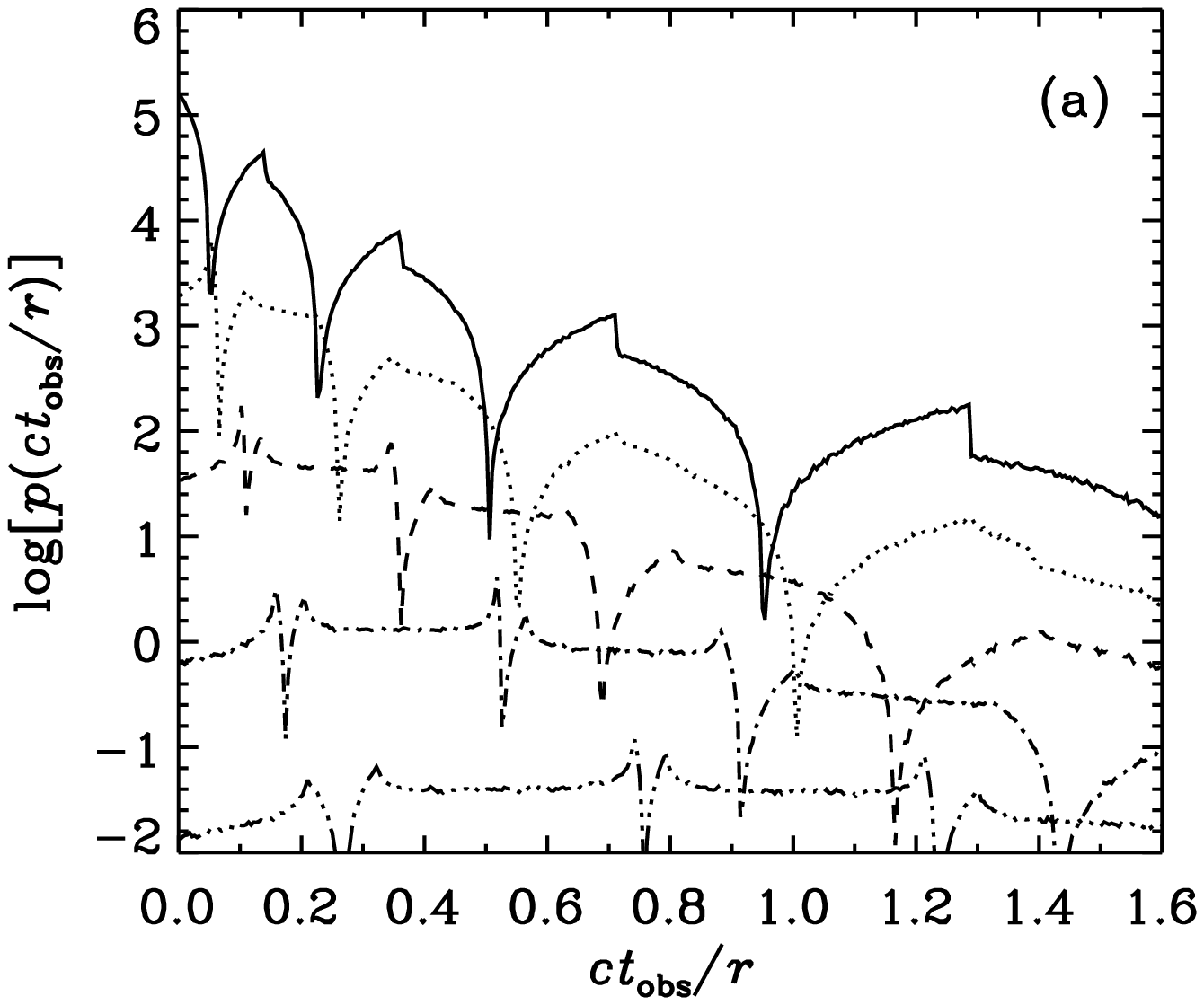, width=100mm}\hspace*{-15mm}
\epsfig{file=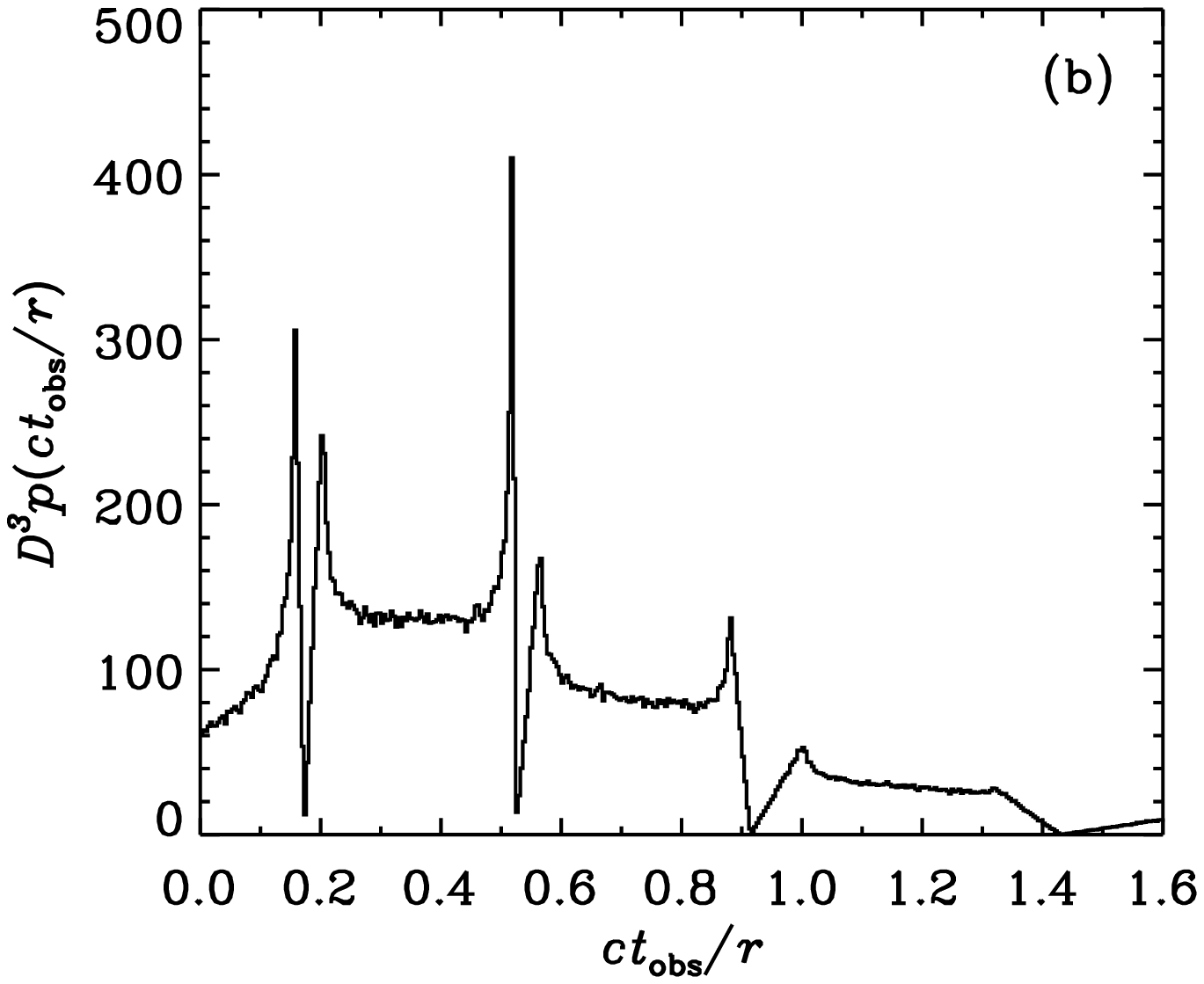, width=100mm}}
\caption{Lightcurve due to excitation of the sequence of
stationary conical conical shocks discussed in the text for an
initial jet Lorentz factor $\Gamma=10$ (a) as seen at viewing
angles $\theta=0^\circ$ (solid curve), $3^\circ$ (dotted curve),
$6^\circ$ (short dashes), $9^\circ$ (dot-dash) and $12^\circ$
(dot-dot-dot-dash) with respect to the jet axis, and (b) as seen
at viewing angles $\theta=9^\circ$ and plotted on a linear
scale.}
\label{LC_of_coneshocks}
\end{figure}

I shall consider next the case of a sequence of stationary
reverse and forward conical shocks.  Although in real AGN jets it may be
possible that the downstream flow is re-accelerated to near the
original value (depending on conditions external to the jet, and
the jet production mechanism), I shall assume that each
successive conical shock causes a reduction in the jet Lorentz
factor by $\Gamma_d/\Gamma_u = 0.801784$.  I shall make an
additional approximation that the divergence/convergence of the
flow caused by the conical shocks can be neglected, and that the
jet flow is always parallel to the jet axis.  The lightcurve due
to a thin, initially flat, disk travelling relativistically along
the jet is then calculated by the Monte Carlo method, in which
``particles'' are placed over the surfaces of all of the cones.
Each cone has the same number of particles, and they would appear
uniformly distributed across the cross section of the jet when
viewed along the jet axis.  A side view showing the location of
these particles is given in Fig.~\ref{LC_of_coneshocks_geom}.  An
initially flat thin disk is launched along the jet with initial
Lorentz factor $\Gamma=10$, and each time a part of the disk
crosses one of the shocks the Lorentz factor of that part of the
disk drops appropriately, distorting the disk.  The time at which
each ``particle'' emits its photon is determined by the time at
which the (distorted) disk reaches the ``particle''.  The resulting
lightcurve for various viewing angles is shown in
Fig.~\ref{LC_of_coneshocks}(a) where the emission is boosted
using the Doppler factor corresponding to the Lorentz factor of
the flow immediately downstream of each shock.  If the Lorentz
factor of the jet decreases at each shock crossing as assumed
here then, because the Doppler factor is also reduced at each
shock crossing, only the first few conical shocks would be prominent
in the lightcurve resulting from a single jet perturbation.  This
is shown in Fig~\ref{LC_of_coneshocks}(b) which gives the
lightcurve from Fig~\ref{LC_of_coneshocks}(a) for
$\theta=9^\circ$ on a linear scale.

As can be seen, the lightcurve is quasi-periodic.  It would be
strictly periodic if the downstream flow velocity were identical
to the upstream flow velocity.  This might be the case if the
re-collimmation of the diverging flow from the conical shocks
results in restoration of the flow Lorentz factor to near the
upstream value, and in this case the flux from successive cycles
would be at about the same level, as observed in Markarian 501
flares.  The two flow velocities would also be roughly the same
if the initial jet Lorentz factor were much higher than
$\Gamma=10$ used in Fig.~\ref{LC_of_coneshocks}(a) such that the
intervals between flares due to a pair of conical shocks was
roughly the same.  However, in this case the flux of successive
flares would diminish as the Doppler factor decreases.  Turning
this argument around, we may have a method of determining the
minimum jet Lorentz factor from observational data, e.g. the
$\sim 23$~day periodicity in the 1997 Markarian 501 data
(Protheroe et al.\ 1998, Hayashida et al.\ 1998) may be used to
put a lower limit on $\Gamma$.

\begin{figure}[hbt]
\begin{center}
\epsfig{file=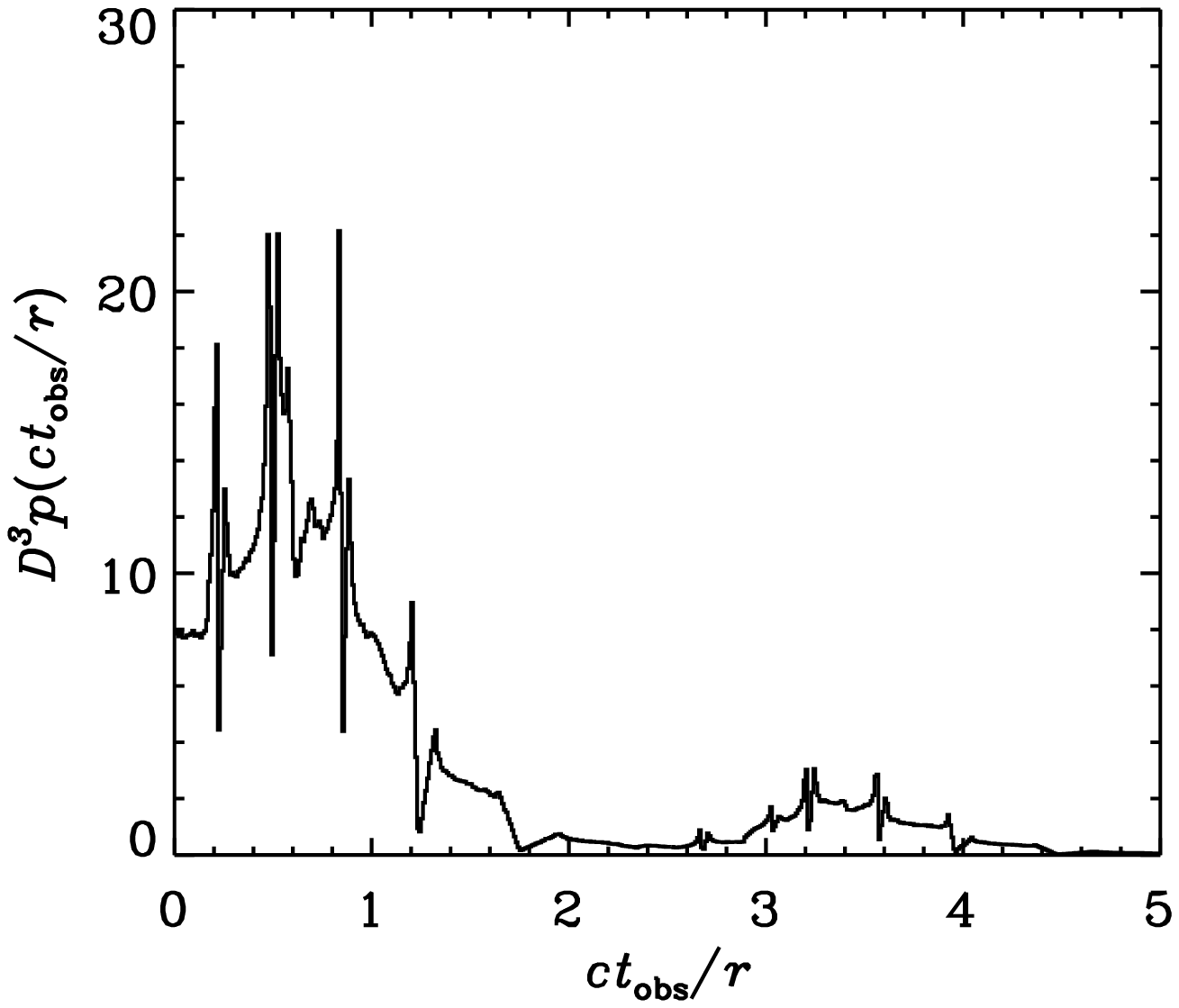, width=100mm}
\caption{Lightcurve due to repeated excitation of the sequence of
stationary conical conical shocks discussed in the text for an
initial jet Lorentz factor $\Gamma=10$ as seen at viewing angles
$\theta=9^\circ$.  The time interval between perturbations
entering the jet is sampled from an exponential distribution with
mean $r/c$, and the strength of the perturbation is also sampled
from an exponential distribution.}
\label{LC_of_coneshocks_superpos}
\end{center}
\end{figure}

Fig.~\ref{LC_of_coneshocks_superpos} shows an example lightcurve
due to multiple jet perturbations.  In this case, I have used an
exponential distribution of times between the injection into the
jet of a density perturbation with a mean interval of 
$r/c$.  I have also used an exponential distribution for the
strength of each perturbation.  The resulting lightcurve looks,
at least qualitatively, as good as any model for the variability
of fluxes from blazars.  Note that even though there is no strict
periodicity resulting from the excitation of a sequence of shocks
in this case, each flare episode has two strong peaks due to (for
this viewing angle) the apex of the first reverse shock cone and
the apex of the first forward shock cone, and in each flare the
separation between these two peaks is identical.  Fourier
analysis would therefore show a strong peak at the frequency
corresponding to the time interval between these two peaks in a
single flare.

\section{Conclusion}

Many factors can influence observed variability time.  The
connection between $\Delta t_{\rm obs}$, Doppler factor and
emission region geometry is non-trivial, and so measuring $\Delta
t_{\rm obs}$ may give, at best, one of the dimensions of the
emission region. Using $R'=c\Delta t_{\rm obs}D$ and $U'_{\rm
phot}=L'/4\pi {R'}^2 c$ may then lead to over-estimation or
under-estimtion of the jet-frame photon energy density by orders
of magnitude.  This is clearly of importance in any AGN model in
which the low energy photons produced in the jet are targets for
interaction of high energy particles or radiation, such as in SSC
models and hadronic blazar models.  Although not discussed in
detail in the present paper, the escape of $\gamma$-rays from the
emission region depends on the optical depth to photon-photon
pair production interactions.  This optical depth can be
uncertain by orders of magnitude in the same way as the photon
energy density, and will also depend on viewing angle.  One must
therefore be careful when using the observation of apparently
unattenuated gamma-rays, and an observed variability time, to
place limits on the Doppler factor.

The uncertainty in the jet-frame photon energy density discussed
in this paper may also have implications for the high brightness
temperature/Compton catastrophe problem of IDV sources.  In this
case, it is the energy density of target photons which limits the
brightness temperature through the competition of inverse-Compton
scattering with synchrotron radiation, and the target photon
energy density may actually be lower than estimated if the emission
region is non-spherical.

If the jet is bent or helical, or has some other favoured
geometry (e.g. conical shocks) cusps in $t'$ vs.\ $t_{\rm obs}$,
and/or a varying Doppler factor may cause narrow peaks in the
observed lightcurve irrespective of other factors.
Distinguishing between these cases from the observed lightcurve
alone is likely to be difficult.  One way of distinguishing
whether a flare is due to (i) an emission region moving around a
bent or helical path, or (ii) a shock is exciting a curved,
conical or helical structure within a jet, is that the in the
first case the flare is caused by a change in viewing angle with
respect to the motion of the emission region leading to a change
in Doppler factor, whereas in the second case the flare is due to
a pile-up in observation times with no change in Doppler factor.
Hence, in case (i) not only will the observed flux increase
during a flare, but the photon energies also increase -- increase
in $(\nu F_\nu)_{\rm peak}$ by factor $x^4$ accompanied by shift
in $\nu_{\rm peak}$ by factor $x$ ($x$ is ratio of final to
initial Doppler factor).  In case (ii), however, since there is
no change in Doppler factor there should be no shift in $\nu_{\rm
peak}$ accompanying an increase in $(\nu F_\nu)_{\rm peak}$.

Distinguishing between the excitation of a conical and a helical
structure by a plane shock would be almost impossible -- note the
qualitative similarities between lightcurves depicted in
Figs.~\ref{agn_tvar_helix}(d) and \ref{LC_of_coneshocks}(b).
Relativistic jet simulations (Bowman 1994, Gomez et al.\ 1997) do
show the presence of conical shocks, and these shocks appear
after a large perturbation (e.g. from 4 to 10 in the simulations
of Gomez et al.\ 1997) in Lorentz factor of the matter entering the
jet through the nozzle.  This can result in a sequence of
quasi-stationary to superluminal reverse and forward conical
shocks extending from the nozzle to the perturbation as it moves
along the jet (Agudo et al.\ 2001).  For the conical shock model
(Salvati et al.\ 1998) discussed here to work, a subsequent
perturbation would need to result in a plane shock, or plane thin
perturbation of some kind, which could travel along the jet and
excite the pre-existing conical shocks.  As far as I am aware,
whether or not this could occur has not been demonstrated.  A
similar uncertainty hangs over whether or not helical jet
structures, which may themselves be shocks with a twisted ribbon
topology, resulting from a perturbation entering the jet through
the nozzle, could subsequently be excited by the passage of a
plane shock or perturbation.  Nevertheless, in both cases, if the
viewing angle is favourable pile-ups in $t_{\rm obs}$, and hence
flares, could occur simply as a result of the motion of the
conical or helical patterns which may themselves be sites of
enhanced emission.  Note that in recent 3D relativistic jet
simulations, the introduction of a 1 percent helical velocity
perturbation at the nozzle results in a helical pattern
propagating along the jet at nearly the beam speed (Aloy et al.\
1999), and that the conical shocks resulting from a perturbation
in jet Lorentz factor can range from being quasi-stationary to
superluminal (Agudo et al.\ 2001).

In conclusion, in models for flaring in AGN in which the emission
comes from a localized region (blob) co-moving with the jet, time
variability is non-trivial to interpret in terms of emission
region geometry and Doppler factor.  A further complication is
that flaring may arise instead due to curved or helical motion of
a blob, even if the emission is constant in the istantaneous rest
frame of the blob.  In this case, apparent flaring is due to the
change in viewing angle, and hence Doppler factor.  Similarly, if
the viewing angle is favourable, relativistic motion of curved or
helical filaments or surfaces can lead to observation of flares.
Excitation of curved or helical jet structures by shocks or
perturbations can also lead to pile-ups in $t_{\rm obs}$, and
hence large apparent increases in flux.  Observations of time
variability in AGN is therefore non-trivial to interpret and may
lead to large systematic errors in estimated jet-frame photon
energy density, Doppler factor and the physical parameters of the
emission region.


\section*{Acknowledgments}

I thank Peter Biermann for helpful discussions and Alina Donea
for a careful reading of the manuscript.  I would like to thank
the anonymous referees for their suggestions which have led to
improvements to the paper.  My research is supported by a grant
from the Australian Research Council and a grant from the
University of Adelaide.

\section*{References}

\reference Agudo, I, Gomez, J.-L., Marti, J.-M., Ibanez, J.-M., Marscher, A.P.,
 Alberdi, A., Aloy, M.-A., Hardee, P.E. 2001, ApJ, 549, L183



\reference Aloy, M.-A., Ibanez, J.-M., Marti, J.M., Gomez, J.-L., Muller,
 E. 1999, ApJ, 523, L125

\reference Bowman, M. 1994, MNRAS, 269, 137

\reference Bednarek, W. \& Protheroe, R.J. 1997, { MNRAS}, { 292}, 646

\reference Bednarek, W. \& Protheroe, R.J. 1999, { MNRAS}, 310, 577

\reference Camenzind, M. 1986, { A\&A}, 156, 136

\reference Catanese, M. et al.\ 1998, { ApJ}, 501, 616 


\reference Chadwick, P.M. et al.\ 1999, ApJ, 513, 161

\reference Despringre, V. \& Fraix-Burnet, D. 1997, { A\&A}, 320, 26 

\reference Gaidos, J.A. et al. 1996, { Nature}, 383, 319

\reference Giebels, B. et al. 2002, ApJ, 571, 763

\reference Gomez, L.L., Marti, J.M., Marscher, A.P., Ibanez, J.M, \& Alberdi, A., 1997, ApJ, 482, L33

\reference Hardee, P.E. 2000, ApJ, 533, 176

\reference Hartman, R.C. et al. 1999, ApJSuppl, 123, 79

\reference Hayashida, N. et al. 1998, ApJ, 504, L71

\reference Kedziora-Chudczer, L., et al.\, 1997, ApJ, 490, L9

\reference Kardashev, N.S., 2000, Astron.\ Reports, 44, 719

\reference Kellerman, K.I., \& Pauliny-Toth, I.I.K., 1969, ApJ, 155, L71

\reference Kifune, T. 2002, PASA, 19, 1-4

\reference Kniffen, D.A. et al. 1993, ApJ, { 411},  133.

\reference Konopelko, A. et al. 1999, Astropart. Phys., 11, 135


\reference Lind, K.R \& Blandford, R.D. 1985, ApJ, 295, 358



\reference Mannheim, K. \& Biermann, P.L., 1989, {A\&A}, {221}, 211

\reference Mannheim, K., Protheroe, R.J. \& Rachen, J., 2001,
Phys.\ Rev.\, D 63, 023003.

\reference M\"{u}cke, A., \& Protheroe, R.J.,  2001, Astropart.\ Phys., 15
121.

\reference M\"{u}cke, A., Protheroe, R.J., Engel, R., Rachen, J. \& Stanev, T., 2002, Astropart.\ Phys., Astropart.\ Phys., submitted.

\reference Marcowith, A., Henri G. \& Pelletier G., 1995, 
	{ MNRAS}, 277, 681.

\reference Mendoza, S. \& Longair, M.S. 2001, MNRAS, 324, 149

\reference von Montigny C. et al., 1995, { ApJ}, { 440}, 525

\reference Mukherjee R. et al., 1997, { ApJ}, { 490}, 116.

\reference Protheroe, R.J. 1997, in Accretion Phenomena and
Related Outflows, IAU Colloq.\ 163. ASP Conf.\ Ser.\ Vol.\ 121,
ed. D.T. Wickramasinghe et al., p.~585

\reference Protheroe, R.J. et al., 1998, Invited, Rapporteur, and
        Highlight Papers of the 25th Int. Cosmic Ray
        Conf. (Durban), eds. M.S. Potgieter et al., pub. World
        Scientific (Singapore), p. 317

\reference Protheroe, R.J. 2002, MNRAS, submitted.

\reference Punch M. et al., 1992, { Nature}, 358, 477

\reference Qian S.-J. et al., 1992, { Chinese A\&A}, 16, 137

\reference Quinn J. et al. 1996, { ApJ}, 456, L83

\reference Rantakyro F. T. et al, 1998, { A\&ASuppl}, 131, 451

\reference Rieger, E.M. \& Mannheim, K. 2000, A\&A, 359, 948

\reference Rosen A., 1990, { ApJ}, 359, 296

\reference Salonen, E., Terasranta, H., Urpo, S., Tiuri, M.,
 Moiseev, I.G., Nesterov, N.S., Valtaoja, E.,
 Haarala, S., Lehto, H., Valtaoja, L., Teerikorpi, P.,
 Valotnen, M. 1987, A\&ASuppl, 70, 409

\reference Salvati, M., Spada, M. \& Pacini, F 1998, ApJ, 495, L19

\reference Schramm K.-J. et al, 1993, { A\&A}, 278, 391

 \reference Spada, M., Salvati, M. \& Pacini, F 1999, ApJ, 511, 136

\reference Steffen W., Zensus, J.A., Krichbaum, T.P.,
 Witzel, A., Qian, S.J., 1995, { A\&A}, 302, 335.

\reference Tateyama C. E. et al, 1998, { ApJ}, 500, 810

\reference Terasranta, H., Tornikoski, M., Valtaoja, E.,
 Urpo, S., Nesterov, N., Lainela, M., Kotilainen, J.,
 Wiren, S., Laine, S., Nilsson, K.,  Valtonen, L. 1992, A\&ASuppl, 94, 121

\reference Urry, C.M. \& Padovani, P. 1995, PASP, 107, 803

\reference Villata, M. \& Raiteri, C.M., 1999, { A\&A}, 347, 30

\reference Wagner, S.J. \& Witzel, A., 1995, { ARA\&A}, 33, 163

\reference Walker, M.A., 1998, MNRAS, 294, 307

\end{document}